\documentclass[aps,prl,reprint,twocolumn,superscriptaddress,amsfonts,amssymb]{revtex4-2}

\usepackage{mathtools}
\usepackage{bm}
\usepackage{graphicx,color}
\usepackage{hyperref}
\usepackage[normalem]{ulem}
\usepackage{cleveref}
\usepackage{microtype}

\newcommand{\Cij}{C_i{}^j}
\newcommand{\hcTil}{\tilde{h}_c}
\newcommand{\etaTil}{\tilde{\eta}}
\newcommand{\myparallel}{{\mkern3mu\vphantom{\perp}\vrule depth 0pt\mkern3mu\vrule depth 0pt\mkern3mu}}

\definecolor{linkblue}{rgb}{0, 0, 0.5}
\hypersetup{
  colorlinks = true, 
  urlcolor = linkblue, 
  linkcolor = linkblue, 
  citecolor = linkblue 
}

\begin{document}

\title{Active nematic flows over curved surfaces}
\author{Samuel Bell*}
\affiliation{Laboratoire Physico-Chimie Curie, UMR 168, Institut Curie, PSL Research University, CNRS, Sorbonne Universit\'{e}, 75005 Paris, France}
\author{Shao-Zhen Lin*}
\affiliation{Aix Marseille Universit\'{e}, Universit\'{e} de Toulon, CNRS, Centre de Physique Th\'{e}orique, Turing Center for Living Systems, Marseille, France}
\author{Jean-Fran\c{c}ois Rupprecht}
\affiliation{Aix Marseille Universit\'{e}, Universit\'{e} de Toulon, CNRS, Centre de Physique Th\'{e}orique, Turing Center for Living Systems, Marseille, France}
\author{Jacques Prost}
\email{Jacques.Prost@curie.fr}
\affiliation{Laboratoire Physico-Chimie Curie, UMR 168, Institut Curie, PSL Research University, CNRS, Sorbonne Universit\'{e}, 75005 Paris, France}
\affiliation{Mechanobiology Institute, National University of Singapore, 117411 Singapore.}
\begin{abstract}
Cell monolayers are a central model system to tissue biophysics. In vivo, epithelial tissues are curved on the scale of microns, and curvature’s role in the onset of spontaneous tissue flows is still not well-understood. Here, we present a hydrodynamic theory for an apical-basal asymmetric active nematic gel on a curved strip. We show that surface curvature qualitatively changes monolayer motion compared to flat space: the resulting flows can be thresholdless, and the transition to motion may change from continuous to discontinuous. Surface curvature, friction and active tractions are all shown to control the flow pattern selected: from simple shear to vortex chains.
\end{abstract}

\date{\today}

\maketitle

Had Gaudi been a developmental biologist rather than an architect, he might have said that there are no flat epithelial tissues in nature. Tissue flows occur in the intrinsically curved environments encountered during morphogenetic processes or inner organs self-renewal, e.g. in the gut, where flows of cells occur along a curvature gradient~\cite{Hannezo2011,Shyer2013}.
Yet, for practical reasons, most \textit{in vitro} studies on collective epithelial tissue flows were performed on flat surfaces. These verified several predictions of active nematic theory, which, applied to tissue, predicts the relations between mechanical stress, flows and cell shape fields, defined through a coarse-grained procedure~\cite{Ramaswamy2003,Marchetti2013,Saw2017,Duclos2018}.

Here, motivated by recent development in 3D micropatterning and live-3D imaging techniques~\cite{XI2022121380}, we investigate theoretically the emergence of sponteneous flows within a covariant active nematic framework for epithelial tissues.  

The effect of curvature on active nematics was explored in several recent studies~\cite{Pearce2019,Pearce2020,Napoli2020} that extended previous equilibrium frameworks~\citep{Biscari2006,Kralj2011,Napoli2012b,Napoli2016}.
These theories are quadratic in the curvature tensor. Yet, epithelial tissues, which have lumen and substrate facing sides -- called apical and basal, respectively, permit a linear coupling to curvature.
Indeed, a body of recent work has shown that the underlying substrate curvature regulates the cellular architecture~\cite{Baptista2019,Callens2020,Harmand2021}, with cells orientation depending on the substrate convexity/concavity, both at the single cell~\cite{Comelles2014,Bade2018,Callens2020} or collective tissue scale level~\cite{Liu2018,Pieuchot2018,Yu2021,Yu2018,Luciano2020}.
Theoretical models that do consider such apico-basal asymmetry~\cite{Dicko2017,Streichan2018, Morris2019} do not, however, address the possibility of a mechanical feedback loop between flows and active stresses.

Here we show that, by affecting the cell orientation, the sign and intensity of the curvature 
alters the nature of the transition to flows in confined active nematic geometry and, in stark contrast to the non-curved case, can lead to thresholdless shear flows at vanishing activity or confinement size.

We first derive an active nematic hydrodynamic framework in the presence of up-down asymmetry within a curved manifold. We then predict new cell-shape and tissue flows patterns within monolayers placed on curved substrates.
We show that there exists a critical curvature value above which a uniform state becomes unstable, distorts, and starts to flow.
The flow mode depends on the magnitude and sign of curvature.
Then, we numerically show the existence of a discontinuous transition in the value of the flow velocity between these previously identified modes, as well as more complicated 2D modes such as vortex chains.
We also highlight the existence of multiple steady state patterns in regions of the phase diagram previously thought to be stable.

We start by describing a fully-developed active nematic phase with a unit-length director field~$\bm{n}$.
Surfaces are characterised by both the metric tensor~$g_{ij}$, and the extrinsic curvature tensor~$\Cij$~\cite{Salbreux2017}.
For convenience, we write an effective free energy~$F=F_0+F_C$, and define the molecular field~$h_i=-\delta F/\delta n^i$, the functional derivative of the total free energy, with ($h_\perp$, $h_\myparallel$) those components respectively perpendicular and parallel to the director field~$\bm{n}$.
$F$ has a part analogous to the Frank free energy of classic liquid crystals: $F_0 = \int dS\sqrt{g}[K_1(\nabla_j n^j)^2/2+K_3(n^j\nabla_j n_i)^2/2-h_{\myparallel}^0n_in^i/2]$, where $K_1$ and $K_3$ are the splay and bend elastic moduli respectively, and $h_\myparallel^0$ is the Lagrange multiplier enforcing the constraint~$\bm{n}^2=1$.
In this paper, we use the one-constant approximation $K_1=K_3=K$~\cite{de1993physics}, motivated by recent observations in epithelial tissues~\cite{Blanch-Mercader2021}.
All $\nabla_i$ terms refer to covariant derivatives.
We use Einstein notation for summation over repeated indices.

\begin{figure}
\centering
\includegraphics[width=\columnwidth]{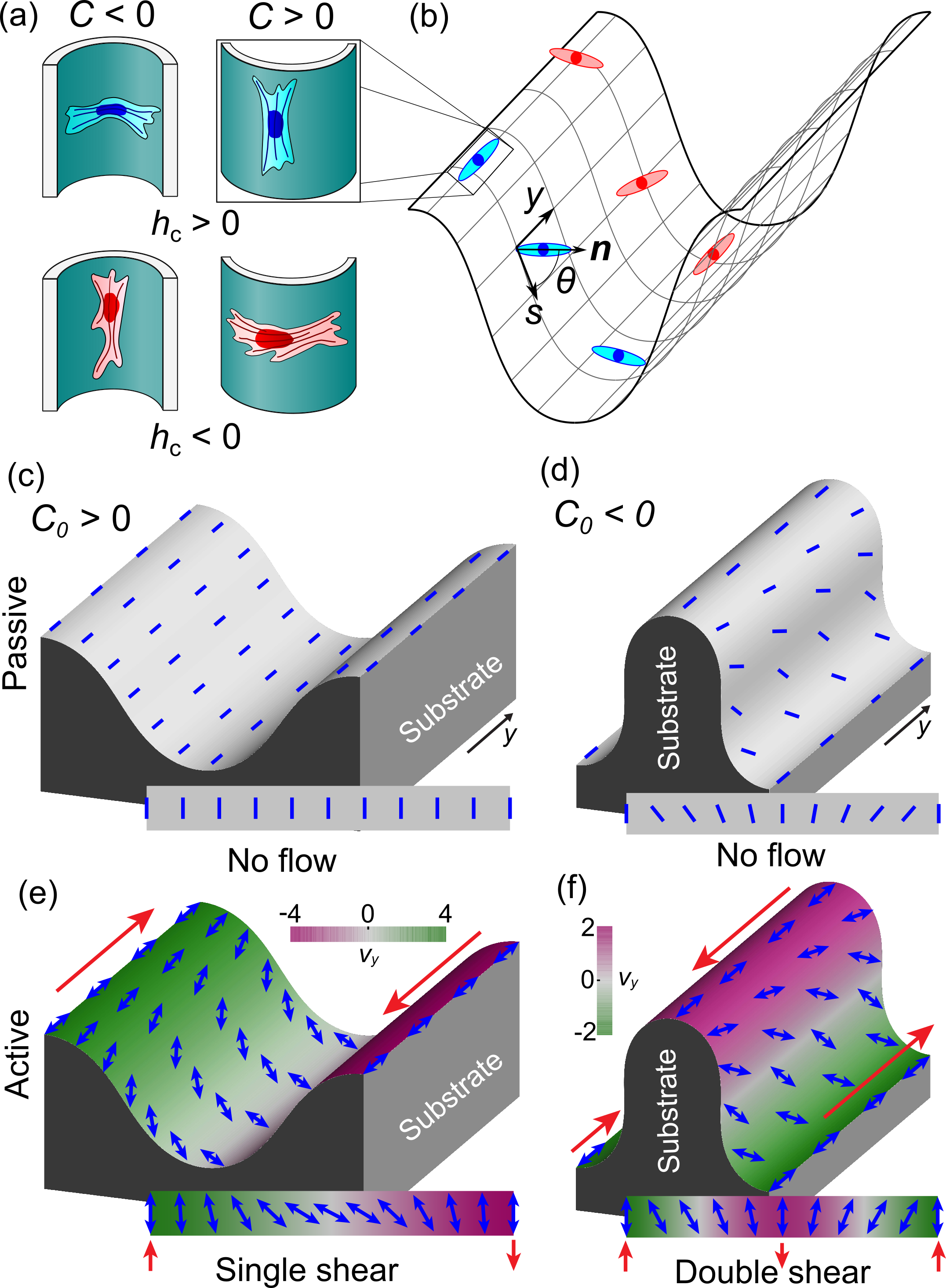}
\caption{(a) Cells with alignment parameter $h_c>0$ in the free energy (blue cells, e.g.\ fibroblasts~\cite{Callens2020}), will align along the direction with the most negative (or least positive) curvature, whereas cells with $h_c<0$ (red cells, e.g.\ MDCK cells~\cite{Callens2020}) will align oppositely.
(b) The surface geometry: a stripe, infinite in $y$, with curvature in the transverse $s$-direction $C_{ss}=C_0\cos 2\pi s/L$, where $L$ is transverse contour length and $C_0>0$, showing single cells aligning according to their alignment parameter.  (c)–(f) Typical cell orientation and velocity patterns of (c) passive uniform pattern, (d) passive non-uniform pattern, (e) active single shear flow pattern and (f) active double shear flow pattern. The color code refers to $v_y$  and the red arrows indicate flow directions. Cartesian plots of the orientation and velocity are given in Fig S1.}
\label{fig:curv}
\end{figure}

Here, we consider a linear curvature free energy,
\begin{equation}
F_C = \frac{1}{2}\int \sqrt{g}\, \mathrm{d}S\, h_c\Cij n^i n_j,
\end{equation}
which is the simplest first order expansion in the curvature field allowed by nematic symmetry.
If $h_c>0$, cells prefer to align parallel to directions of greatest negative curvature, and perpendicular to directions of greatest positive curvature, and vice versa for $h_c<0$, see Fig.~\ref{fig:curv}(a).

The monolayer's velocity with respect to a fixed substrate, $\bm{v}$, enters into tensors for the strain rate $u_i{}^j=(\nabla_i v^j+\nabla^j v_i-\nabla_k v^k g_i{}^j)/2$, and vorticity $\omega_i{}^j=(\nabla_i v^j-\nabla^j v_i)/2$.
For simplicity, we assume incompressibility, $\nabla\cdot \bm{v}=0$.
The stress constitutive equation reads $\sigma_i{}^j=-P\delta_i{}^j+\tilde{\sigma}_i{}^j+\sigma^A{}_i{}^j+\sigma^a{}_i{}^j$  for active gels~\cite{Marchetti2013}, where: $P$ collects up all the isotropic stresses; $\tilde{\sigma}_i{}^j=2\eta u_i{}^j+\frac{\nu}{2}(n_i h^j +n^j h_i-n_k h^k g_i{}^j)$, where $\eta$ is the kinematic viscosity, and $\nu$ is the shear alignment coefficient; $\sigma^A{}_i{}^j=-(n_i h^j-n^j h_i)/2$; $\sigma^a{}_i{}^j=-\zeta(n_i n^j-g_i{}^j/2)$ is the active stress~\cite{Prost2015}.
The equation for the evolution of the director field reads
\begin{equation}\label{eq:Pol}
\frac{D n_i}{D t}=\frac{h_i}{\gamma}-\nu  u_i{}^j n_j+\nu_c C_i{}^j n_j , 
\end{equation}
where $Dn_i/Dt = \partial n_i/\partial t+v^j\nabla_j n_i+\omega_i{}^j n_j$ is the corotational derivative of the director field $n_i$; $\nu$ is the shear alignment coefficient.
The last term of Eq.~\eqref{eq:Pol} is an active term different from the terms generated by the coupling in the effective free energy.
We explore the role of this term in the Supplemental Material~\cite{SI}, Fig. S3; we assume $\nu_c=0$ in the rest of the paper.

The force balance between internal stresses and momentum exchanges with the substrate reads
\begin{equation}
\nabla^j\sigma_j{}_i=\xi v_i+\lambda_b n^j\nabla_j n_i+\lambda_s n_i\nabla_j n^j,  \label{eq:stressequation}
\end{equation}
where $\xi$ is a substrate friction coefficient; $\lambda_b$ (resp. $\lambda_s$) is an active bend (resp. splay) coefficient, expressing that an active nematic can specifically extract momentum from the substrate for bend or splay conformations~\cite{Maitra2018}~\footnote{Here we will neglect possible curvature-gradient active forces, which would lead to an extra term $\lambda_C\nabla_j\Cij$ in Eq.~\eqref{eq:stressequation}.}.
Recent studies have shown that the magnitude of curvature regulates the migration mode of monolayers confined to a tube~\cite{Xi2017}; for simplicity, we do not include these contributions in the current study. The sum $\lambda_s +  \lambda_b$ combination of these terms simply amounts to a shift in the overall value of the active stress $\zeta$.
Their difference, $\lambda_s - \lambda_b$, has no bulk counterpart in classical active gel theory~\cite{Prost2015}.
It has been recently considered, but only within unbound domains~\cite{Maitra2018}.

We focus here on the geometry of an infinite strip of width $L$, Fig~\ref{fig:curv}a).
For in-plane curvilinear coordinates $(s,y)$, $C_{ss}(s) = C_0 \cos ks$ is the only non-zero component of the curvature tensor, with $C_0$ being the curvature magnitude and $k = 2 \pi /L$. In the spirit of Voituriez et al.~\cite{Voituriez2005}, we first assume that the flow and orientation patterns are invariant along the $y$-direction. Coupled with the incompressibility condition, this \hbox{$y$-invariance} of the system implies that $v_s=0$.
Thus, the force balance in the $s$-direction defines the pressure $P$.
We need only consider one off-diagonal stress component: $\sigma_{sy}$.
We write $\sigma_{sy}$, the force balance, and Eq.~\eqref{eq:Pol}, in terms of the molecular field components $(h_\perp,h_\myparallel)$ and the director field angle $\theta$ (see Fig.~\ref{fig:curv}(b)), such that $\bm{n}=(\cos\theta,\sin\theta)$ (SM \cite{SI}, Sec. I).
The $y$-invariant system admits two uniform non-moving solutions at steady state: $\theta=0$ and $\theta=\pi/2$, depending on the boundary conditions.
Here, we focus on the homogeneous alignment condition (observed in experiments~\cite{Duclos2018}) which permits one uniform solution, $\theta=\pi/2$ (the director field aligned parallel to the boundaries).

\begin{figure}[t!]
\centering
\includegraphics[width=\columnwidth]{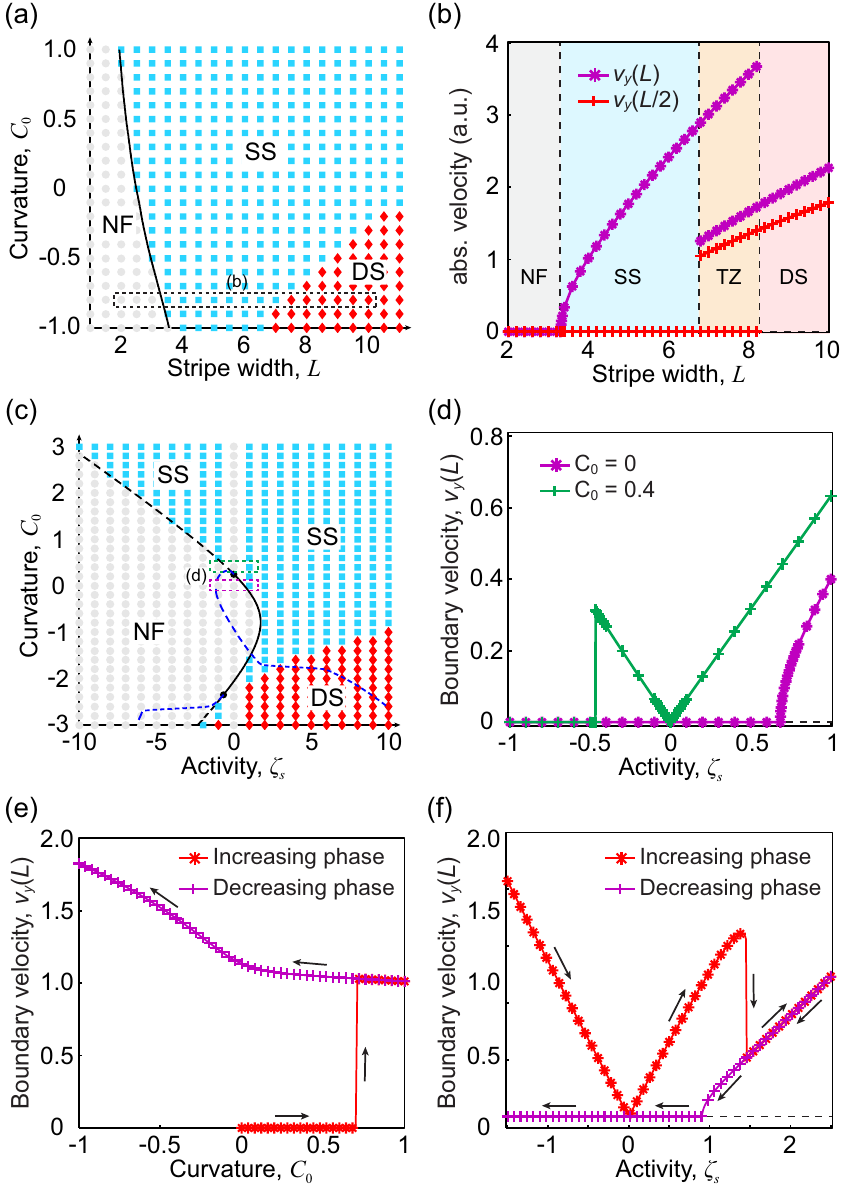}
\caption{Flow patterns and transitions for $\xi=0$ obtained from numerical simulations (SM \cite{SI}, Sec. I). 
(a) $C_0$--$L$ flow pattern diagram for $\zeta_s=4$: no-flow (grey, NF); single shear (blue, SS); and double shear (red, DS) and analytical prediction Eq. (\ref{eq:lambda1eq0}) of the continuous NF--SS transition (black solid curve). 
(b) Constant $C_0=-0.8$ cut through (a) (indicated by dotted box) showing continuous NF--SS transition, and discontinuous SS--DS transition.
(c) $\zeta_s$--$C_0$ flow pattern diagram and analytical prediction Eq. (\ref{eq:lambda1eq0}) of the NF--SS transition (solid: continuous; dashed: discontinuous). 
The two black circular spots separate the continuous transition and the discontinuous transition.
The blue dashed curve represents the discontinuous SS--NF transition while increasing (or decreasing) $\zeta_s$ from a SS pattern (see Fig. S6). 
(d) Constant $C_0$ cuts ($C_0=0$ and $C_0=0.4$, the dashed boxes in (c)) showing a discontinuous NF--SS transition for the curved case, with thresholdless flows around zero activity.
(e) Pattern selection upon varying $C_0$ quasi-statically (see SI Sec. I.D) with $\zeta_s = -1.5$, where the black arrows indicate the route. See also Movie S1. 
(f) Pattern selection upon varying $\zeta_s$ quasi-statically (see SI Sec. I.D) with $C_0 = -1.6$. See also Movie S2. 
Parameters other than $\xi$ in SM Table 1.}
\label{fig:nofric}
\end{figure}

We first analyse the linear stability of a perturbation to the uniform state, $\delta\theta=\theta-\pi/2$, with boundary conditions $\delta\theta(0)=\delta\theta(L)=0$.
In the zero friction $\xi=0$ limit and with stress-free boundary conditions $\sigma_{sy}(0)=\sigma_{sy}(L)=0$, the force balance in the $y$-direction can be integrated to give $\sigma_{sy}=-\lambda_s\delta\theta$.
Replacing $\sigma_{sy}$ in the stress equation shows that the active splay renormalizes the contractility, $\zeta\to\zeta+\lambda_s\equiv\zeta_s$.
Similarly, the active torque renormalizes the curvature free energy coefficient, $h_c\to h_c-\etaTil\nu_c\equiv \hcTil$, where $\etaTil=4\eta\gamma/(\gamma(\nu+1)^2+4\eta)$ is an effective viscosity.
Introducing adimensional units $\tilde{s} = k s/2$, $\tau = (Kk^2/4\etaTil)t$, the dynamical equation for the perturbation $\delta\theta$ can be written (SM \cite{SI}, Sec. II):
\begin{equation}
\frac{\partial\delta\theta}{\partial\tau} = 
\left(-\frac{2\zeta_s(\nu+1)\etaTil}{ K k^2\eta}
- \frac{4\hcTil C_0}{Kk^2}\cos2\tilde{s}\right)\delta\theta
+ \frac{\partial^2\delta\theta}{\partial\tilde{s}^2},
\end{equation}
or, $\dot{\delta\theta}=\mathcal{L}(\tilde{s})\delta\theta$, where $\mathcal{L}(\tilde{s})$ is the operator of the Mathieu equation, $\mathcal{L}(\tilde{s})=(a-2q\cos 2\tilde{s})+\partial_{\tilde{s}}^2$~\cite{dlmf}.
The eigenvalues and eigenfunctions of this operator, $\mathcal{L}(\tilde{s})\phi_m=\lambda_m\phi_m$~\cite{dlmf}, determine the stability of the system; since $\dot{\phi}_m=\lambda_m\phi_m$, any $\lambda_m>0$ implies that the uniform solution is unstable.
For $\delta\theta(0)=\delta\theta(L)=0$, the eigenvalues are related to the odd characteristic numbers $b_m(q)$ of the Mathieu equation~\cite{dlmf}.
The lowest eigenfunction, $\phi_1$, is a single shear (SS) pattern, akin to Fig.~\ref{fig:curv}(e).
The second eigenfunction, $\phi_2$, is a double shear (DS) pattern as in Fig.~\ref{fig:curv}(f), where two shear bands are stitched together in the centre of the strip. The condition that the lowest mode vanishes,
\begin{equation}\label{eq:lambda1eq0}
\lambda_1(q) = -\frac{\zeta_s(\nu+1)\etaTil L^2}{2\pi^2\eta K }+b_1(q) = 0,
\end{equation}
is a criteria for the transition from the non-flowing (NF) state to a single shear state. We found an excellent agreement between Eq. (\ref{eq:lambda1eq0}) and numerical simulations throughout the $(C_0,\zeta_s,L)$ phase space, Figs. \ref{fig:nofric} and~\cite{SI}.

For small values of curvature $q = 2\hcTil C_0/Kk^2 \ll 1$, the expansion $b_1(q)\sim 1 + q$ yields the following curvature threshold for flow:
\begin{equation}
\frac{\hcTil C_{0,cr}}{2K} = \left(\frac{\pi}{L}\right)^2 +\frac{\zeta_s(\nu+1)\etaTil}{2\eta K}.
\label{eq:linthresh}
\end{equation}
Such approximation of the critical curvature $C_{0, cr}$ agrees well with simulations, Fig. S2(a). When $C_0=0$ this expression reduces to that of the active Fredericksz transition found in~\cite{Voituriez2005}. As visible in Eq. (\ref{eq:linthresh}), the curvature field induces a new length scale, $(2K/\hcTil C_0)^{1/2}$, which pits the cost of non-alignment with the curvature against the cost of director field deformations. 
As in flat space~\cite{Duclos2018,Voituriez2005}, increasing the strip width $L$ can induce a continuous NF--SS transition, Figs. \ref{fig:nofric}(a,b).
For fixed $L$, the NF--SS threshold is a non-monotonic function of the curvature, see Fig.~\ref{fig:nofric}(c); such behavior, which contrasts with the linear relation Eq.~\eqref{eq:linthresh}, is a consequence of the higher order terms in the $b_1(q)$ expansion at larger $q$.

Our simulations also reveal transitions to higher order shear modes, each differentiated by their velocities in the centre, $v_y(L/2)$, and edge, $v_y(L)$, of the strip: $v_y(L/2)=v_y(L)=0$ for NF; $v_y(L/2)=0$, $v_y(L)\neq 0$ for SS; and $v_y(L/2),v_y(L)\neq 0$ for DS. In particular, the DS pattern emerge for $\hcTil C_0 <0$, see Fig.~\ref{fig:nofric}(a)~and~(c).
When $\hcTil C_0 >0$, cells in the centre of the strip prefer to align towards $\theta=0$; the central portion is biased towards a large perturbation from $\theta=\pi/2$, favoring the SS pattern. When $\hcTil C_0<0$, this is reversed.
The cells in the centre of the strip prefer to align towards $\theta=\pi/2$, favoring the DS pattern.
There is no direct NF--DS transition for $\xi=0$, see Figs. \ref{fig:nofric}(b). 
The SS--DS transition is discontinuous, with a transition zone of metastability.

We find that the NF--SS transition is not always continuous (see Fig.~\ref{fig:nofric}(d) and Fig. S5).
At two tricritical points (the black dots in Fig.~\ref{fig:nofric}(c)), the transition changes from continuous (solid black line) to discontinuous (dashed black line). The upper tricritical point occurs right at $\zeta_s = 0$, see Fig.~\ref{fig:nofric}(c) and Fig. S5.
Close to the lower tricritical point, the growth rates of the SS and DS modes are nearly degenerate, $\lambda_1\approx\lambda_2$.
A higher-order analysis of a mixed state in this region shows that couplings between the shear modes can effectively reverse the sign of the third order term of the amplitude equation (SM \cite{SI}, Sec. II), a typical hallmark of a tricritical point, offering excellent agreement with the numerical simulations. 

We then explored the regimes of hysteresis along the discontinuous NF--SS transition upon performing quasi-static variation of the curvature. Indeed, the NF--SS transition and the SS--NF transitions occur at different values of $C_0$, see Fig.~\ref{fig:nofric}(e) and Movie S1. 
The lines of these transitions are known as the spinodal lines.
The boundary of linear stability of the NF state, $\lambda_1 = 0$ gives one of these spinodal lines for each tricritical point.
To find the other spinodals (the blue dotted lines in Fig.~\ref{fig:nofric}(c)), we prepare a single shear state and then transform the system quasi-statically (varying $C_0$ or $\zeta_s$) until the shear state lost absolute stability (SM \cite{SI}, Sec. I; Fig. \ref{fig:nofric}(f) and Fig. S6; Movie S2). 
We note that the upper blue spinodal line crosses $C_0 = 0$, implying that a single shear state can be at least metastable in flat space for both signs of $\zeta_s$. Such a behavior is a departure from the established work on the continuous active shear transition in flat stripes~\cite{Voituriez2005}. 

We find that the $\zeta_s=0$ line cuts through the domain of stability of the single shear patterns. There, curvature destabilises the uniform director field pattern (see Eq.~\eqref{eq:linthresh} and Figs. \ref{fig:curv}(c,d)).
The resulting orientation gradients drive motion even for arbitrarily small active stresses and the flow velocity scales linearly as $v_y\propto\zeta_s$, see Fig.~\ref{fig:nofric}(d). Similar thresholdless active flows are found in \cite{Green2017} as a consequence of the constraints imposed by the anchoring condition.

\begin{figure}[t!]
\centering
\includegraphics[width=\columnwidth]{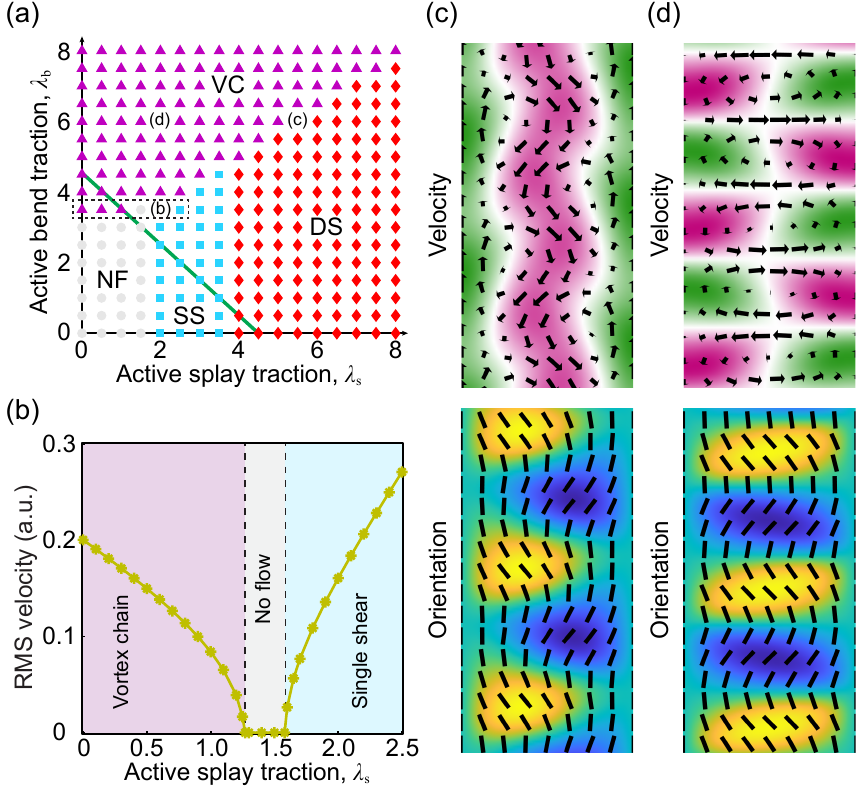}
\caption{The effect of active bend traction $\lambda_b$. Results shown here were obtained from full 2D simulations. (a) Phase diagram of the flow patterns (stable (light grey); single shear (blue); double shear (red) and vortex chain (magenta)) with respect to the active splay traction $\lambda_s$ and the active bend traction $\lambda_b$. (b) Phase transition regulated by the active splay traction $\lambda_s$, which corresponds to the horizontal dash box in (a) with $\lambda_b = 3.5$. (c, d) The perturbed double shear flow pattern (c) and the vortex chain pattern (d), which correspond to the data set shown in (a). In the velocity map, the color code represents $v_y$ and arrows denote velocity vectors; in the orientation map, the color code refers to $\theta$ and lines for orientation directors. Parameters: $\xi = 0.6$, $L=8$, and $C_0 = 0$.}
\label{fig:2D}
\end{figure}

When friction is non-zero, there is no tidy equation in $\delta\theta$. We expand in Fourier space and find that the spatial dependence of the curvature field leads to mode coupling (SM \cite{SI}, Sec. II). 
We find the critical curvature $C_{0,cr}$ for small curvature values:
\begin{equation}
\frac{h_\xi C_{0,cr}}{2K} = \left(\frac{\pi}{L}\right)^2\left(1+\frac{(\nu+1)\zeta_s}{2K[\xi/\gamma+(\pi/L)^2(\eta/\tilde{\eta})]}\right) , \label{eq:CcritPos}
\end{equation}
where we define a renormalized curvature-coupling coefficient: $h_\xi=h_c-\nu_c(\xi+\eta(\pi/L)^2)/(\xi/\gamma+\pi^2\eta/ L^2\tilde{\eta})$.
Equation \eqref{eq:CcritPos} shows excellent agreement with numerical simulations (Fig. S2(b)) and matches Eq. \eqref{eq:linthresh} for a vanishing substrate friction $\xi=0$.
Higher absolute curvatures are needed to initiate flows with larger friction, see Fig. S2(b). In contrast to the zero friction case, we find a direct and continuous transition from the non-flowing to double shear patterns, see Fig. S8.

To check the assumption of $y$-invariance, we perform two-dimensional simulations, SM Sec. I \cite{SI}. For $\lambda_b = 0$, the assumption holds very well (Fig. \ref{fig:2D}a and Fig. S9). However, for large enough active bend traction $\lambda_b > 0$, longitudinal instabilities appear (Fig. \ref{fig:2D}c), eventually leading to a vortex chain pattern (Fig. \ref{fig:2D}d and Fig. S11)~\cite{Doostmohammadi2016}.
The vortex chain state shares a phase boundary with the NF, SS and DS patterns (Fig.~\ref{fig:2D}(a)).
Recalling that the sum of the active bend and splay correspond to a bulk renormalization of the contractility, $\zeta\to\zeta+(\lambda_s+\lambda_b)/2$, we can consider systems of equal contractility by taking diagonal cuts of Fig.~\ref{fig:2D}(a) such as the green line.
For a given value of the contractility, we may observe each of the four flow patterns, depending on the ratio between the active splay and bend tractions.

\textbf{Perspectives} In this work, we have travelled beyond the existing paradigm of continuous transition to flows in confined active nematics~\cite{Voituriez2005}.
The introduction of a curvature field allows for tight control of a wider variety of flow structures than simple shear states, each with unique biological significance (e.g. a double shear flow permits net transport for weak anchoring boundary conditions).
Further, by suppressing the threshold to motion, curvature vastly increases the range of tissue parameters that allow for flows.
Finally, our findings of discontinuous transitions and hysteresis offer new perspectives on tissue dynamics: even transient mechanical perturbations, such as a brief shock to the tissue, may lead to long-term changes from a quiescent state to a spontaneously flowing one.
This concept that living matter, epithelial tissues included, could exist with hair triggers to motion has the potential to transform our understanding of morphogenesis, cancer spreading, or biofilm growth.

\textbf{Acknowledgements}
S. B. and J. P. are funded by the Human Frontiers in Science Program (HFSP RGP0038/2018) and acknowledge useful discussions with other members of the project.
J.-F. R. and S-Z. L. are funded by the Investissements d’Avenir French Government program managed by the French National Research Agency (ANR-16-CONV- 0001 and ANR-17-CE13-0032 COVFEFE), the A*MIDEX Excellence Initiative of Aix-Marseille University.

%

\end{document}


\title{Supplementary Material \\ Active nematic flows over curved surfaces}
\author{Samuel Bell}
\affiliation{Laboratoire Physico-Chimie Curie, UMR 168, Institut Curie, PSL Research University, CNRS, Sorbonne Universit\'{e}, 75005 Paris, France}
\author{Shao-Zhen Lin}

\author{Jean-Fran\c{c}ois Rupprecht}
\affiliation{Aix Marseille Universit\'{e}, Universit\'{e} de Toulon, CNRS, Centre de Physique Th\'{e}orique, Turing Center for Living Systems, Marseille, France}
\author{Jacques Prost}
\affiliation{Laboratoire Physico-Chimie Curie, UMR 168, Institut Curie, PSL Research University, CNRS, Sorbonne Universit\'{e}, 75005 Paris, France}
\affiliation{Mechanobiology Institute, National University of Singapore, 117411 Singapore.}

\date{\today}
\maketitle

\makeatletter
\renewcommand\@dotsep{10000}
\makeatother

\tableofcontents
\listoffigures
\listoftables

\section{Numerical simulations}

\subsection{Governing equations}

In the main text, we have given the general governing equations (Eqs. (1-3)) for active nematic gels on curved surfaces and focus on a simple curved geometry, i.e., curved infinite stripes, as shown in Fig. 1(b). We here give the detailed governing equations and boundary conditions for active nematic gels on such curved geometry. 

For a curved stripe geometry as shown in Fig. 1(b), the extrinsic curvature tensor reads, 
\begin{equation}
C_{ss}=C_0 \cos \left( \frac{2\pi s}{L} \right) = C_0 \cos(ks), \quad C_{yy} = C_{sy} = C_{ys} = 0 , 
\end{equation}
where $k = 2\pi/L$. The strain rate tensor $E_{\alpha\beta}$ and the vorticity tensor $\Omega_{\alpha\beta}$, respectively, read, 
\begin{equation}
E_{ss} = \frac{\partial v_s}{\partial s}, \quad E_{yy} = \frac{\partial v_y}{\partial y} = -E_{ss}, \quad E_{sy} = \frac{1}{2}\left( \frac{\partial v_y}{\partial s} + \frac{\partial v_s}{\partial y} \right), \quad E_{ys} = \frac{1}{2}\left( \frac{\partial v_y}{\partial s} + \frac{\partial v_s}{\partial y} \right) = E_{sy}
\end{equation}
and
\begin{equation}
\Omega_{ss} = \Omega_{yy} = 0, \quad \Omega_{sy} = \frac{1}{2}\left( \frac{\partial v_y}{\partial s} - \frac{\partial v_s}{\partial y} \right), \quad \Omega_{ys} = \frac{1}{2}\left( \frac{\partial v_s}{\partial y} - \frac{\partial v_y}{\partial s} \right) = -\Omega_{sy}
\end{equation}
The evolution equation of the director field reads 
\begin{align}
\frac{\partial n_s}{\partial t} &= -\left( v_s\frac{\partial n_s}{\partial s} + v_y\frac{\partial n_s}{\partial y} \right) - \left( \Omega_{ss} n_s + \Omega_{sy} n_y \right) -\nu \left( E_{ss} n_s + E_{sy} n_y \right) + \nu_c C_{ss} n_s + \frac{1}{\gamma }h_s , \\
\frac{\partial n_y}{\partial t} &= -\left( v_s\frac{\partial n_y}{\partial s} + v_y\frac{\partial n_y}{\partial y} \right) - \left( \Omega_{ys} n_s + \Omega_{yy} n_y \right) - \nu \left( E_{ys} n_s + E_{yy}n_y \right) + \frac{1}{\gamma }h_y . 
\end{align}
Since $n_s = \cos\theta$ and $n_y = \sin\theta$, we have, 
\begin{align}
-\sin\theta \frac{\partial\theta }{\partial t} &= \sin\theta \left( v_s\frac{\partial\theta }{\partial s} + v_y\frac{\partial\theta }{\partial y} \right) - \left( \Omega _{ss}\cos\theta + \Omega_{sy}\sin\theta  \right) - \nu\left( E_{ss}\cos\theta + E_{sy}\sin\theta  \right) + \nu_c C_{ss}\cos\theta + \frac{1}{\gamma}h_s , \label{eq_n_evolution_s} \\
\cos\theta\frac{\partial\theta}{\partial t} &= -\cos\theta\left(v_s\frac{\partial\theta }{\partial s} + v_y\frac{\partial\theta }{\partial y} \right) - \left(\Omega_{ys}\cos\theta + \Omega_{yy}\sin\theta\right) - \nu\left( E_{ys}\cos\theta +E_{yy}\sin\theta  \right)+\frac{1}{\gamma}h_y . \label{eq_n_evolution_y}
\end{align}
Taking $-\sin\theta\times\text{Eq.}~\eqref{eq_n_evolution_s}+\cos\theta\times\text{Eq.}~\eqref{eq_n_evolution_y}$ we obtain the evolution equation of the orientation field as, 
\begin{equation}
\frac{\partial\theta}{\partial t} = -\left( v_s\frac{\partial\theta}{\partial s} + v_y\frac{\partial\theta}{\partial y} \right) + \Omega_{sy} + \nu\left(E_{ss}\sin 2\theta - E_{sy}\cos 2\theta\right)-\frac{1}{2}\nu_c C_{ss}\sin 2\theta + \frac{1}{\gamma}h_\bot , \label{eq_theta_evolution}
\end{equation}
where
\begin{equation}
h_\bot = -\dfrac{\delta F}{\delta \theta} = K\left(\frac{\partial^2\theta}{\partial s^2} + \frac{\partial^2\theta}{\partial y^2} \right) + \frac{1}{2}h_c C_{ss}\sin 2\theta , 
\end{equation}
is the perpendicular component of the molecular field; here we have assumed equal splay and bend elasticity, i.e., $K_1 = K_3 = K$. For completeness, we further take $\cos\theta\times \text{Eq.}~\eqref{eq_n_evolution_s}+\sin\theta\times\text{Eq.}~\eqref{eq_n_evolution_y}$ and get the parallel component of the molecular field as,
\begin{equation}
\frac{1}{\gamma}h_\parallel = \nu\left( E_{ss}\cos 2\theta + E_{sy}\sin 2\theta  \right) - \nu_c C_{ss}\cos^2\theta.
\end{equation}
The force balance equation reads, 
\begin{align}
\frac{\partial\sigma_{ss}}{\partial s} + \frac{\partial\sigma_{ys}}{\partial y} - \xi v_s + T_s &= 0 , \label{eq_ForceBalance_s} \\
\frac{\partial\sigma_{sy}}{\partial s} + \frac{\partial\sigma_{yy}}{\partial y} - \xi v_y + T_y &= 0 , \label{eq_ForceBalance_y}
\end{align}
where $T_s$ and $T_y$ are the active tractions, stemming from the splay and bend deformations, 
\begin{equation}
\begin{aligned}
  & T_s = -\lambda_s\cos\theta\left(-\sin\theta\frac{\partial\theta}{\partial s} + \cos\theta\frac{\partial\theta}{\partial y}\right) + \lambda_b\sin\theta\left(\cos\theta\frac{\partial\theta}{\partial s} + \sin\theta\frac{\partial\theta}{\partial y}\right), \\ 
 & T_y = -\lambda_s\sin\theta\left( -\sin\theta\frac{\partial\theta}{\partial s} + \cos\theta\frac{\partial\theta}{\partial y} \right) - \lambda_b\cos\theta\left(\cos\theta\frac{\partial\theta}{\partial s} + \sin\theta \frac{\partial\theta}{\partial y}\right). \\ 
\end{aligned}
\end{equation}
The stress components are
\begin{equation}
\begin{split}
  & \sigma_{ss} = -P + 2\eta\frac{\partial v_s}{\partial s} + \mathrm{X}, \\ 
 & \sigma_{yy} = -P - 2\eta\frac{\partial v_s}{\partial s} - \mathrm{X}, \\ 
 & \sigma_{sy} = \eta\left(\frac{\partial v_y}{\partial s} + \frac{\partial v_s}{\partial y}\right) + \mathrm{Y} + \mathrm{Z}, \\ 
 & \sigma_{ys} = \eta\left(\frac{\partial v_y}{\partial s} + \frac{\partial v_s}{\partial y}\right) + \mathrm{Y} - \mathrm{Z}, \\ 
\end{split} \label{eq_stress}
\end{equation}
where 
\begin{align}
  \mathrm{X} &= \frac{1}{2}\nu\left( 
  h_\parallel \cos 2\theta - h_\bot \sin 2\theta  \right) - \frac{1}{2}\zeta\cos 2\theta, \label{eq:X}  \\ 
  \mathrm{Y} &= \frac{1}{2}\nu\left(h_\parallel \sin 2\theta + h_\bot\cos 2\theta \right) - \frac{1}{2}\zeta\sin 2\theta, \label{eq:Y}  \\ 
  \mathrm{Z} &= -\frac{1}{2}h_\bot, \label{eq:Z} 
\end{align}
Substituting Eq.~\eqref{eq_stress} into Eqs.~\eqref{eq_ForceBalance_s} and \eqref{eq_ForceBalance_y}, we obtain, 
\begin{align}
-\xi v_s + \eta\left(\frac{\partial^2 v_s}{\partial s^2} + \frac{\partial^2 v_s}{\partial y^2}\right) &= \frac{\partial P}{\partial s} - \frac{\partial\mathrm{X}}{\partial s} - \frac{\partial\mathrm{Y}}{\partial y} + \frac{\partial\mathrm{Z}}{\partial y} - T_s, \label{eq_ForceBalance_s_1} \\
-\xi v_y + \eta\left(\frac{\partial^2 v_y}{\partial s^2} + \frac{\partial^2 v_y}{\partial y^2}\right) &= \frac{\partial P}{\partial y} + \frac{\partial\mathrm{X}}{\partial y} - \frac{\partial\mathrm{Y}}{\partial s} - \frac{\partial\mathrm{Z}}{\partial s} - T_y. \label{eq_ForceBalance_y_1}
\end{align}
In addition, the incompressibility condition reads, 
\begin{equation}
\frac{\partial v_s}{\partial s} + \frac{\partial v_y}{\partial y} = 0. \label{eq_Incompressibility}
\end{equation}

\textit{Boundary conditions} -- We consider the following boundary conditions. For the orientation field $\theta\left( \mathbf{r},t \right)$, we assume homogeneous anchoring boundary conditions at $s = 0$ and $s = L$, that is, 
\begin{equation}
\left.\theta\right|_{s=0} = \left.\theta\right|_{s=L} = \frac{\pi}{2}. \label{eq_BC_theta}
\end{equation}
For the velocity and stress fields, we assume no lateral flow and no shear stress at the lateral "wall" ($s = 0$ and $s = L$); such condition reads
\begin{equation}
\left. v_s\right|_{s=0} = \left. v_s\right|_{s=L} = 0, \label{eq_BC_NoLateralFlow}
\end{equation}
and 
\begin{equation}
\left.\sigma_{sy}\right|_{s=0} = \left.\sigma_{sy}\right|_{s=L} = 0 . \label{eq_BC_NoShearStress}
\end{equation}
Substituting the expression of stress Eq.~\eqref{eq_stress} into \eqref{eq_BC_NoShearStress}, we can re-express the stress boundary condition Eq.~\eqref{eq_BC_NoShearStress} in terms of the velocity gradient as
\begin{equation}
\begin{aligned}
  & \left.\frac{\partial v_y}{\partial s} \right|_{s=0} = -\frac{1}{\eta}\left.\left(\text{Y} + \text{Z} \right)\right|_{s=0} = \frac{\left(\nu + 1 \right)K}{2\eta}\left.\frac{\partial^2\theta }{\partial s^2} \right|_{s=0}, \\ 
 & \left.\frac{\partial v_y}{\partial s} \right|_{s=L} = -\frac{1}{\eta}\left.\left (\text{Y} + \text{Z}\right)\right|_{s=L} = \frac{\left(\nu + 1\right)K}{2\eta}\left. \frac{\partial^2\theta}{\partial s^2} \right|_{s=L}, 
\end{aligned} \label{eq_BC_NoShearStress_1}
\end{equation}
in terms of the function defined in Eqs. (\ref{eq:Y}) and (\ref{eq:Z}).

In summary, we have derived the governing equations of the orientation field $\theta(\mathbf{r},t)$ (Eq. \eqref{eq_theta_evolution}) and the flow field $\mathbf{v}(\mathbf{r},t)$ (Eqs. \eqref{eq_ForceBalance_s_1} and \eqref{eq_ForceBalance_y_1}). These equations together with the boundary conditions (Eqs. \eqref{eq_BC_theta}, \eqref{eq_BC_NoLateralFlow} and \eqref{eq_BC_NoShearStress_1}) determine the stability and dynamics of active nematic gels on curved infinite stripes. 

\subsection{In one-dimension}
Let us assume here an uniform system along the $y$-axis, i.e., $\theta = \theta(s,t)$ and $\mathbf{v} = \mathbf{v}(s,t)$. Then the incompressibility condition Eq. \eqref{eq_Incompressibility} together with the no lateral flow boundary condition Eq. \eqref{eq_BC_NoLateralFlow} implies that $v_s = 0$. The evolution equation of cell orientation field $\theta(s,t)$ simplifies into
\begin{equation}
\frac{\partial\theta}{\partial t} = \frac{1}{2}\left(1 - \nu\cos 2\theta  \right)\frac{\partial v_y}{\partial s} + \frac{1}{\gamma}h_\bot - \frac{1}{2}\nu _c C_{ss}\sin 2\theta, \label{eq_theta}
\end{equation}
where $h_\bot=K\dfrac{\partial^2\theta}{\partial s^2}+\frac{1}{2}h_c C_{ss}\sin 2\theta$. The force balance equation simplifies to: 
\begin{equation}
-\xi v_y + \eta\frac{\partial^2 v_y}{\partial s^2} = -\frac{\partial\mathrm{Y}}{\partial s} - \frac{\partial\mathrm{Z}}{\partial s} - T_y, 
\end{equation}
in terms of the function defined in Eqs.~\eqref{eq:X},~\eqref{eq:Y},~\eqref{eq:Z}, which then leads to
\begin{equation}
-\xi v_y + \gamma\nu^2 \sin 2\theta \cos 2\theta \frac{\partial\theta}{\partial s}\frac{\partial v_y}{\partial s} + \left(\eta + \frac{1}{4}\gamma \nu^2 \sin^2 2\theta\right)\frac{\partial^2 v_y}{\partial s^2} = \frac{\partial\mathrm{W}}{\partial s}, \label{eq_ForceBalance}
\end{equation}
where 
\begin{equation}
\mathrm{W} = \frac{1}{2}\left(\lambda_b - \lambda_s\right)\theta + \frac{1}{4}\left(2\zeta + \lambda_b + \lambda_s\right)\sin 2\theta + \frac{1}{4}\gamma\nu\nu_c C_{ss}\sin 2\theta \left(1 + \cos2\theta\right) + \frac{1}{2}h_\bot\left(1 - \nu\cos2\theta\right),
\end{equation}
is a function of the cell orientation angle $\theta$. 

We use a finite difference method to numerically solve the governing equations \eqref{eq_theta} and \eqref{eq_ForceBalance} with the boundary conditions \eqref{eq_BC_theta} and \eqref{eq_BC_NoShearStress_1}. In brief, we firstly update the cell orientation angle $\theta \left(t \right)\to \theta \left( t+\Delta t \right)$ using the forward Euler scheme. Then we solve the velocity $v_y\left( t+\Delta t \right)$ based on the updated cell orientation angle $\theta \left( t+\Delta t \right)$ using a central finite difference scheme~\cite{Lapidus_2011_book}. 

If not otherwise stated, in our simulations, we set our default parameter values as detailed in Table \ref{table_s1}. For the passive case, we further set $\nu_c = 0$ and $\zeta = \lambda_s = \lambda_b = 0$. The space and time steps are typically set as $\Delta s = 1/16$ and $\Delta t = 0.001$, respectively. 

\subsection{In two-dimension}

Here we give the numerical schemes for solving the fully two-dimensional governing equations for active nematic gels on a curved infinite stripe geometry, as shown in Fig. 1(b). For such curved geometry, we simulate active nematic gels in a square domain $[0, L] \times [0, L_y]$ in the $(s,y)$ plane. To mimic the infinity of the curved stripe, we set $L_y \gg L$ and apply periodic boundary conditions along the $y$-axis, that is, 
\begin{equation}
\theta(s,y,t) = \theta(s,y+L_y,t), \quad v_s(s,y,t) = v_s(s,y+L_y,t), \quad v_y(s,y,t) = v_y(s,y+L_y,t)
\end{equation}

We use the finite difference method to solve the orientation field $\theta(\mathbf{r},t)$ and the velocity field $\mathbf{v}(\mathbf{r},t)$. Specifically, we firstly update the orientation field $\theta(\mathbf{r},t) \rightarrow \theta(\mathbf{r},t+\Delta t)$ using a forward Euler scheme, 
\begin{equation}
\theta\left(\mathbf{r},t + \Delta t \right) = \theta\left(\mathbf{r},t\right) + g_\theta\left( \mathbf{r},t\right)\Delta t, 
\end{equation}
where 
\begin{equation}
g_\theta = -\left(v_s\frac{\partial\theta }{\partial s} + v_y\frac{\partial\theta }{\partial y}\right) + \Omega_{sy} + \nu\left(E_{ss}\sin2\theta - E_{sy}\cos2\theta\right) - \frac{1}{2}\nu_c C_{ss}\sin2\theta + \frac{1}{\gamma}h_\bot
\end{equation}

Based on the updated orientation field $\theta(\mathbf{r},t+\Delta t)$, we next employ the stream function method to solve the force balance equations \eqref{eq_ForceBalance_s_1} and \eqref{eq_ForceBalance_y_1}. In the stream function method, the incompressibility condition \eqref{eq_Incompressibility} is naturally satisfied. We introduce the stream function $\psi$ as, 
\begin{equation}
v_s = \frac{\partial\psi}{\partial y}, \quad v_y = -\frac{\partial\psi}{\partial s}, \label{eq:streamfunction}
\end{equation}
The force balance equations \eqref{eq_ForceBalance_s_1} and \eqref{eq_ForceBalance_y_1} can be re-expressed as: 
\begin{align}
-\xi \frac{\partial\psi}{\partial y} - \eta\frac{\partial\omega}{\partial y} - \frac{\partial P}{\partial s} + \frac{\partial\mathrm{X}}{\partial s} + \frac{\partial\mathrm{Y}}{\partial y} - \frac{\partial\mathrm{Z}}{\partial y} + T_s &= 0, \label{eq_ForceBalance_s_2} \\
\xi \frac{\partial\psi}{\partial s} + \eta\frac{\partial\omega}{\partial s} - \frac{\partial P}{\partial y} - \frac{\partial\mathrm{X}}{\partial y} + \frac{\partial\mathrm{Y}}{\partial s} + \frac{\partial\mathrm{Z}}{\partial s} + T_y &= 0, \label{eq_ForceBalance_y_2} 
\end{align}
where 
\begin{equation}
\omega = -\left( \frac{\partial^2 \psi}{\partial s^2} + \frac{\partial^2 \psi}{\partial y^2}\right) \label{eq_omega_psi}
\end{equation}
is the vorticity. 

To eliminate the pressure term, we apply the differential operator $\partial / \partial y$ to Eq.~\eqref{eq_ForceBalance_s_2} and the differential operator $\partial / \partial s$ to Eq.~\eqref{eq_ForceBalance_y_2}. Subtracting the resulting two equations, we obtain that
\begin{equation}
\xi\omega - \eta\nabla^2 \omega = -2\frac{\partial^2\mathrm{X}}{\partial s\partial y} + \frac{\partial^2\mathrm{Y}}{\partial s^2} - \frac{\partial^2\mathrm{Y}}{\partial y^2} + \frac{\partial^2\mathrm{Z}}{\partial s^2} + \frac{\partial^2\mathrm{Z}}{\partial y^2} + \frac{\partial T_y}{\partial s} - \frac{\partial T_s}{\partial y}. \label{eq_ForceBalance_omega}
\end{equation}

Using the stream function $\psi$ defined in Eq. (\ref{eq:streamfunction}), the no-lateral flow boundary conditions Eq. \eqref{eq_BC_NoLateralFlow} reads
\begin{equation}
\left.\frac{\partial\psi}{\partial y}\right|_{s=0} = \left.\frac{\partial\psi}{\partial y}\right|_{s=L} = 0 , 
\end{equation}
which leads to 
\begin{equation}
\left.\psi\right|_{s=0} = 0, \quad \left.\psi\right|_{s=L} = C, \label{eq_BC_NoLateralFlow_1}
\end{equation}
where the constant $C$ is determined by the force balance equation in the $y$-axis, as to be shown in the following. In addition, the no-shear stress boundary conditions Eq. \eqref{eq_BC_NoShearStress_1} reads
\begin{equation}
\left.\omega\right|_{s=0} = \frac{\left(\nu + 1\right)K}{2\eta}\left.\frac{\partial^2\theta}{\partial s^2} \right|_{s=0}, \quad \left.\omega\right|_{s=L} = \frac{\left(\nu + 1\right)K}{2\eta}\left.\frac{\partial^2\theta}{\partial s^2}\right|_{s=L}. \label{eq_BC_NoShearStress_2}
\end{equation}

To determine the constant $C$ in Eq. \eqref{eq_BC_NoLateralFlow_1}, we consider the force balance equation in the $y$-axis (Eq. \eqref{eq_ForceBalance_y_2}) and integrate it in the whole curved stripe domain, that is,
\begin{equation}
\int_{L_s\times L_y}\left(
\xi\frac{\partial\psi }{\partial s} + 
\eta\frac{\partial\omega}{\partial s} - 
\frac{\partial P}{\partial y} - 
\frac{\partial\mathrm{X}}{\partial y} + 
\frac{\partial \mathrm{Y}}{\partial s} + 
\frac{\partial\mathrm{Z}}{\partial s} + T_y
\right)\mathrm{d}s\mathrm{d}y = 0,
\end{equation}
which, together with the boundary condition Eq. \eqref{eq_BC_NoShearStress_2}, leads to
\begin{equation}
\left.\psi\right|_{s=L} - \left.\psi\right|_{s=0} = 0.
\end{equation}
Combining the later relation with the no lateral flow boundary condition Eq. \eqref{eq_BC_NoLateralFlow_1}, we therefore obtain that 
\begin{equation}
\left.\psi\right|_{s=0} = \left.\psi\right|_{s=L} = 0. \label{eq_BC_NoLateralFlow_2}
\end{equation}

In summary, we have derived the governing equations of the vorticity field $\omega(\mathbf{r},t)$ and the stream function $\psi(\mathbf{r},t)$, as well as the boundary conditions Eqs. \eqref{eq_BC_NoShearStress_2} and \eqref{eq_BC_NoLateralFlow_2}, which can be solved using the finite difference method~\cite{Lapidus_2011_book}. In details, we first solve the vorticity field $\omega(\mathbf{r},t + \Delta t)$ from Eq. \eqref{eq_ForceBalance_omega} together with the boundary condition Eq. \eqref{eq_BC_NoShearStress_2}, basing on the updated orientation field $\theta(\mathbf{r}, t+\Delta t)$. Subsequently, we solve the stream function $\psi(\mathbf{r},t+\Delta t)$ from Eq. \eqref{eq_omega_psi} together with the boundary condition Eq. \eqref{eq_BC_NoLateralFlow_2}, basing on the updated vorticity field $\omega (\mathbf{r}, t+ \Delta t)$.

\textit{Discretization} -- Space and time steps are typically set as $\Delta s = \Delta y = 1/8$ and $\Delta t = 10^{-4}$, respectively.

\textit{Initial condition} -- We set the initial orientation field as small perturbations around the uniform equilibrium state, $\theta(\mathbf{r},t=0) = {\pi}/{2} + \varepsilon\chi(\mathbf{r})$ where $\varepsilon \ll 1$ and $\chi(\mathbf{r})$ is a Gaussian noise with a zero mean and a unit variance. The initial velocity field $\mathbf{v}(\mathbf{r},t=0)$ is calculated from the initial orientation field $\theta(\mathbf{r},t=0)$ by solving the force balance equations.

\subsection{Quenching simulation scheme} \label{sec:QuenchingSimulation}

\textit{Curvature $C_0$} -- To further investigate the discontinuous NF-SS transition as shown in Fig. 2(c) in the main text, we perform quenching simulations of the curvature $C_0$. In such quenching simulations, we start from $\left( \lambda_s , C_0 \right) = \left( \lambda_s , 0 \right)$ with $\lambda_s < 0$ deeply in the non-flowing state (e.g., $\lambda_s = -1.5$ in Fig. 2(c)). We set the initial orientation pattern $\theta(\mathbf{r},t=0)$ as uniform ($\theta = \pi / 2$) but with small perturbations. We increase the curvature $C_0$ quasi-statically as below: (i) we relax the system for enough time (typically with simulation steps $N_{\rm relax} = 100,000$) to obtain the steady state for the current curvature $C_0^{(i)}$; (ii) once we get the steady state, we increase the curvature to $C_0^{(i+1)} = C_0^{(i)} + \Delta C_0$ with $\Delta C_0 = 0.01$. We repeat above steps (i) and (ii) until in the single shear flow regime. Afterwards, we decrease the curvature $C_0$ quasi-statically via a similar scheme. 

\textit{Activity $\lambda_s$ (or $\zeta$)} -- We have tried two kinds of activity parameters ($\lambda_s$ and $\zeta$). The quenching simulation schemes are similar for $\lambda_s$ and $\zeta$. For simplicity, we here describe the quenching simulation scheme of $\lambda_s$. In such quenching simulations, we do not start from a uniform pattern but instead start from an initial pattern obtained from the decreasing phase of the quenching simulation of the curvature $C_0$. We increase the activity parameter $\lambda_s$ quasi-statically as below: (i) we relax the system for enough time (typically with simulation steps $N_{\rm relax} = 100,000$) to obtain the steady state for the current activity value $\lambda_s^{(i)}$; (ii) once we get the steady state, we increase the activity $\lambda_s^{(i+1)} = \lambda_s^{(i)} + \Delta \lambda_s$ with $\Delta \lambda_s = 0.01$. We repeat above steps (i) and (ii) until in the single shear flow regime. Afterwards, we decrease the activity $\lambda_s$ quasi-statically via a similar scheme.

\section{Analytical results}

In this section, we focus on the infinite stripe geometry and assume as invariant along the $y$-direction. In such case, the governing equations can be simplified as: 
\begin{gather}
\frac{\partial\theta}{\partial t} = \frac{h_\perp}{\gamma} - \frac{\partial_s v_y}{2}\left(\nu\cos 2\theta - 1\right) - \frac{\nu_c C_{ss}\sin 2\theta}{2}, \label{eq:Pol}\\
\partial_s \sigma_{sy} = \xi v_y + (\lambda_b\cos^2\theta - \lambda_s\sin^2\theta)\partial_s\theta, \label{eq:StrMom}
\end{gather}
where
\begin{gather}
\sigma_{sy} = \eta\partial_s v_y-\frac{\zeta\sin 2\theta}{2}+\frac{\nu}{2}(h_\myparallel\sin2\theta+h_\perp\cos2\theta)-\frac{h_\perp}{2} , \label{eq:StrSig} \\
\frac{h_\myparallel}{\gamma} = \frac{\nu\partial_s v_y\sin 2\theta}{2}-\nu_c C_{ss}\cos^2\theta , \label{eq:hparallel} \\
h_\bot = K\frac{\partial^2\theta}{\partial s^2} + \frac{1}{2}h_c C_{ss}\sin2\theta. \label{eq:hbot}
\end{gather}

We take strong anchoring boundary conditions for the director field:
\begin{equation}
\theta(0) = \theta(L) = \frac{\pi}{2}
\end{equation}
and also impose that the system be stress-free at the boundary:
\begin{equation}
\sigma_{sy}(0) = \sigma_{sy}(L) = 0
\end{equation}
In the following sections, we obtain analytical results in the no-friction case for linear and non-linear stability, before we discuss the linear stability of the friction case.

\subsection{No-friction case}

\subsubsection{Linear stability analysis}
To perform a linear stability analysis, we analyse a perturbation $\theta=\pi/2+\delta\theta$ (such that $\delta\theta(0) = \delta\theta(L) = 0$) to linear order in $\delta\theta$.
The linearised system corresponding to Eq. (\ref{eq:Pol},\ref{eq:StrMom},\ref{eq:StrSig}) reads
\begin{equation}
\dot{\delta\theta} = \frac{h_\perp}{\gamma}+\frac{\partial_s v_y(\nu+1)}{2}+\nu_c C_{ss}\delta\theta, \qquad \partial_s \sigma_{sy}=\xi v_y -\lambda_s\partial_s\delta\theta, \qquad \sigma_{sy} =\eta\partial_s v_y +\zeta_s \delta\theta-\frac{h_\perp(\nu+1)}{2}. \label{eq:LinSys}
\end{equation}
For zero friction, $\xi=0$, as described in the main text, the system~\eqref{eq:LinSys} collapses down into a single Mathieu equation governing the dynamics of the direction field perturbation $\delta\theta(s)$:
\begin{equation}
\frac{\partial\delta\theta}{\partial t} = -\left(\frac{\zeta_s(\nu+1)}{	2\eta}+\frac{\tilde{h}_c C_0}{\tilde{\eta}}\cos ks\right)\delta\theta+\frac{K}{\tilde{\eta}}\frac{\partial^2\delta\theta}{\partial s^2}, \label{eq:linearoperator}
\end{equation}
where $\tilde{\eta}=4\eta\gamma/(4\eta+\gamma(\nu+1)^2)$, $\zeta_s=\zeta+\lambda_s$ and $\tilde{h}_c = h_c-\tilde{\eta}\nu_c$. We transform the equation into its standard form by introducing adimensional units $\tilde{s}=ks/2$, $\tau = (K k^2/4\tilde{\eta})t$:
\begin{equation}
\frac{\partial \delta\theta}{\partial\tau}=-\left(\frac{2\zeta_s(\nu+1)\tilde{\eta}}{ K k^2\eta}+\frac{4\tilde{h}_cC_0}{K k^2}\cos 2\tilde{s}\right)\delta\theta+\frac{\partial^2\delta\theta}{\partial \tilde{s}^2}.
\end{equation}
The advantage of this is that we can use standard results for the Mathieu equation -- normally in such an analysis, as we shall see for the case with friction, we explore the stability of Fourier modes, which is complicated here by the position dependence of the curvature field. Instead, we consider the stability equation as an eigenvalue problem, searching for the (dimensionless, for the moment) eigenvalues $\tilde{\lambda}_m$ and eigenfunctions $\phi_m(\tilde{s})$ which satisfy the equation
\begin{equation}
-\left(\frac{2\zeta_s(\nu+1)\tilde{\eta}}{ K k^2\eta}+\frac{4\tilde{h}_cC_0}{K k^2}\cos 2\tilde{s}\right)\phi_m(\tilde{s})+\frac{\partial^2\phi_m(\tilde{s})}{\partial\tilde{s}^2}=\tilde{\lambda}_m\phi_m(\tilde{s}),
\end{equation}
according to the boundary conditions $\phi_m(0)=\phi_m(\pi)=0$. We will write this explicitly in the standard form for the Mathieu equation:
\begin{equation}
(a_m-2q\cos 2\tilde{s})\phi_m(\tilde{s})+\frac{\partial^2\phi_m(\tilde{s})}{\partial \tilde{s}^2}=0,
\end{equation}
where 
\begin{equation}
a_m=-2\zeta_s(\nu+1)\tilde{\eta}/ (K k^2\eta)-\tilde{\lambda}_m, \qquad \mathrm{and} \qquad q=2\tilde{h}_cC_0/Kk^2. \label{eq:qdefinition}
\end{equation}
The boundary conditions, requiring that $\phi_m(\pi)=\phi_m(0)=0$, are fulfilled by the odd eigenfunctions of the Mathieu equation, $\mathrm{se}_m(\tilde{s},q)$. This limits the values of $a_m$ to the odd characteristic values of the Mathieu equations, $b_m(q)$.
Thus, the dimensionless eigenvalues $\tilde{\lambda}_m$ may be written:
\begin{equation}
\tilde{\lambda}_m(q) = -\frac{2\zeta_s(\nu+1)\tilde{\eta}}{Kk^2\eta}-b_m(q).
\label{eq:lambda}
\end{equation}
To find the NF--SS phase boundary, it suffices to plot the threshold expression $\lambda_1(q) = 0$, as in Fig.~\ref{figS_AmpEq}(a), which, as shown in the main text, agrees excellently to the phase boundary found by the non-linear simulations.
\subsubsection{Existence of a tricritical point}
In \textit{dimensional} units, the first two eigenvalues $\lambda_{1}$ and $\lambda_{2}$ can be written
\begin{align}
\lambda_1(q) = -\frac{\zeta_s(\nu+1)}{2\eta}-\frac{\pi^2 K}{\tilde{\eta} L^2}b_1(q), \qquad  \mathrm{and} \qquad 
\lambda_2(q) = -\frac{\zeta_s(\nu+1)}{2\eta}-\frac{\pi^2 K}{\tilde{\eta} L^2}b_2(q),
\end{align}
with the corresponding odd eigenfunctions of the Mathieu equation:
\begin{align}
\phi_1(s) = \mathrm{se}_1(\pi s/L,q), \qquad  \mathrm{and} \qquad 
\phi_2(s) = \mathrm{se}_2(\pi s/L,q)
\end{align}
where $q$ is defined in Eq. (\ref{eq:qdefinition}).
Although it always true that $\lambda_1>\lambda_2$, for $q<0$, these eigenvalues become close to degenerate, see Fig.~\ref{figS_AmpEq}(a).
To understand the phase behaviour better around this point, it is necessary to expand the system to third order.
First, we look at the force balance equation, given strong anchoring and zero-stress boundary conditions, still in the regime of zero friction, $\xi=0$:
\begin{equation}
\begin{split}
\partial_s \sigma_{sy} &= \left(\lambda_b\cos^2\theta-\lambda_s \sin^2\theta \right)\partial_s\theta = \left(\frac{\lambda_b(1+\cos2\theta)}{2}-\frac{\lambda_s(1-\cos 2\theta)}{2}\right)\partial_s\theta, \\
\implies \sigma_{sy} &= \frac{\lambda_b-\lambda_s}{2}\left(\theta-\frac{\pi}{2}\right)+\frac{(\lambda_b+\lambda_s)\sin2\theta}{4} \equiv \frac{\Delta\lambda}{2}\left(\theta-\frac{\pi}{2}\right)+\frac{\bar{\lambda}\sin2\theta}{2}
\end{split}
\end{equation}
Substituting for $\sigma_{sy}$ and $h_{\parallel}$ in Eq.~\eqref{eq:StrSig}:
\begin{equation}
\frac{\Delta\lambda}{2}\left(\theta-\frac{\pi}{2}\right)+\frac{\bar{\lambda}\sin2\theta}{2} = \left(\eta+\frac{\gamma\nu^2\sin^2 2\theta}{4}\right)\partial_s v_y - \frac{\zeta\sin2\theta}{2} +\frac{h_\perp(\nu\cos2\theta-1)}{2}-\frac{\gamma\nu\nu_c C_{ss} \cos^2\theta\sin2\theta}{2}.
\end{equation}
Rearranging this and substituting for $\partial_s v_y$ in Eq.~\eqref{eq:Pol}, we obtain a single equation for the director field dynamics, which may be written in terms of a perturbation to the uniform state $\phi = \theta-\pi/2$:
\begin{align}
\dot{\phi} =  h_\perp\left(\frac{1}{\gamma}+\frac{(\nu\cos 2\phi+1)^2}{4\eta+\gamma\nu^2\sin^2 2\phi}\right)+\frac{(\nu\cos2\phi+1)}{4\eta+\gamma\nu^2\sin^2 2\phi}\left(\Delta\lambda\phi-(\zeta+\bar{\lambda})\sin 2\phi \right)& \notag\\
+\frac{\nu_c C_{ss}\sin2\phi}{2}\left(1-\frac{2\gamma\nu\sin^2\phi(\nu\cos 2\phi+1)}{4\eta+\gamma\nu^2\sin^2 2\phi}\right)&,
\end{align}
where $h_\perp = K \partial_s^2\phi-h_c C_{ss}\sin 2\phi/2$. Expanding to third order in $\phi$, one can collect the terms of the dynamical equation as
\begin{equation}\label{eq:DynEqfull}
\dot{\phi}=\mathcal{L}(\phi, \partial_s^2\phi)+(\mathcal{H}C_{ss}+\mathcal{Z})\phi^3+\mathcal{K}K\phi^2\partial_s^2\phi,
\end{equation}
where 
\begin{enumerate}
    \item the first term in the r.h.s to Eq. (\ref{eq:DynEqfull}) is 
a linear operator (already encountered in the linear stability analysis, Eq. (\ref{eq:linearoperator})).
\begin{align}
\mathcal{L}(\phi,\partial_s^2\phi)&= \frac{1}{\tilde{\eta}}\left(K\partial_s^2\phi-\tilde{h}_c C_{ss}\phi\right)-\frac{\zeta_s(\nu+1)}{2\eta}\phi, \end{align}
\item  the second term in $\phi^3$ in the r.h.s to Eq. (\ref{eq:DynEqfull}) is separated into a curvature-dependent part, proportional to the quantity
\begin{align}
\mathcal{H} &= \frac{\nu(\nu+1)}{\eta}\left(1+\frac{\gamma\nu(\nu+1)}{4\eta}\right)h_c-\frac{\gamma\nu(\nu+1)}{2\eta}\nu_c+\frac{2\tilde{h}_c}{3\tilde{\eta}}, \label{eq:Hdef}
\end{align}
and an activity-dependent part, proportional to the quantity
\begin{align}
\mathcal{Z} &= \frac{\nu+1}{3\eta}(\zeta+\bar{\lambda})+\frac{\nu}{\eta}\left(1+\frac{\gamma\nu(\nu+1)}{2\eta}\right)\zeta_s, \label{eq:Zdef}
\end{align}
\item the last term in the r.h.s to Eq. (\ref{eq:DynEqfull}) is expressed in terms of the quantity
\begin{align}
\mathcal{K}&= -\frac{\nu(\nu+1)}{\eta}\left(1+\frac{\gamma\nu(\nu+1)}{4\eta}\right). \label{eq:Kdef}
\end{align}
\end{enumerate}
\vskip0.5cm
\paragraph*{Amplitude equation} To find the amplitude equations, we take a mixed state solution of the form $\phi = \Psi_1 \phi_1(s)+\Psi_2\phi_2(s)$, and substitute into Eq.~\eqref{eq:DynEqfull}:
\begin{equation}\label{eq:mixSt}
\dot{\Psi}_1\phi_1+\dot{\Psi}_2\phi_2 = \lambda_1(q)\Psi_1\phi_1+\lambda_2(q)\Psi_2\phi_2+(\mathcal{H}C_{ss}+\mathcal{Z})(\Psi_1\phi_1+\Psi_2\phi_2)^3
+\mathcal{K}(\Psi_1\phi_1+\Psi_2\phi_2)^2K\partial_s^2(\Psi_1\phi_1+\Psi_2\phi_2)
\end{equation}
We neglect spatial variation of the amplitudes $(\Psi_1,\Psi_2)$, and then substitute the second derivative term with the linear eigenvalue operator:
\begin{equation}
\begin{split}
K\partial_s^2\phi_m &= \tilde{\eta}\mathcal{L}(\phi,\partial_s^2\phi) + \frac{\zeta_s(\nu+1)\tilde{\eta}}{2\eta}\phi+\tilde{h}_c C_{ss}\phi \\
&= \left(\tilde{\eta}\lambda_m(q)+\frac{\zeta_s(\nu+1)\tilde{\eta}}{2\eta}+\tilde{h}_c C_{ss}\right)\phi \\
&= \left(-\frac{\zeta_s(\nu+1)\tilde{\eta}}{2\eta}-\frac{\pi^2 K}{ L^2}b_m(q)+\frac{\zeta_s(\nu+1)\tilde{\eta}}{2\eta}+\tilde{h}_c C_{ss}\right)\phi \\
&= \left(-\frac{\pi^2 K}{ L^2}b_m(q)+\tilde{h}_c C_{ss}\right)\phi
\end{split}
\end{equation}
where we have used the definition of Eq. (\ref{eq:lambda}). Substituting the latter equatio into Eq.~\eqref{eq:mixSt}, we find that
\begin{align}
\dot{\Psi}_1\phi_1+\dot{\Psi}_2\phi_2 = & \ \lambda_1(q)\Psi_1\phi_1+\lambda_2(q)\Psi_2\phi_2+(\mathcal{H}C_{ss}+\mathcal{Z})(\Psi_1\phi_1+\Psi_2\phi_2)^3 \notag \\
& +\mathcal{K}(\Psi_1\phi_1+\Psi_2\phi_2)^2\Psi_1\left(-\frac{\pi^2 K}{ L^2}b_1(q)+\tilde{h}_c C_{ss}\right)\phi_1 \notag \\
& +\mathcal{K}(\Psi_1\phi_1+\Psi_2\phi_2)^2\Psi_2\left(-\frac{\pi^2 K}{ L^2}b_2(q)+\tilde{h}_c C_{ss}\right)\phi_2
\end{align}
Then, we perform an inner product to isolate each amplitude equation for $\Psi_1$ and $\Psi_2$. We define two inner product quantities $\alpha_{i,j}$ and $\beta_{i,j}$, depending on the presence or not of the curvature field in the inner product:
\begin{align}
\alpha_{i,j} &= \frac{2}{L}\int_0^L\phi_1^i\phi_2^j\;\mathrm{d}s \\
\beta_{i,j} &= \frac{2}{L}\int_0^L C_{ss}(s)\phi_1^i\phi_2^j\;\mathrm{d}s.
\end{align}
The eigenfunctions are orthogonal and normalized:
\begin{equation}
\frac{2}{L}\int_0^L \phi_i\phi_j\;\mathrm{d}s = \delta_{ij}.
\end{equation}
Since $\phi_1$ and $C_{ss}$ are symmetric across the strip while  $\phi_2$ is antisymmetric, all inner products with ($i$, $j = 4-i$) odd are zero:
\begin{equation}
\alpha_{1,3} = \alpha_{3,1} = \beta_{1,3} = \beta_{3,1} = 0.
\end{equation}
Performing the inner products, we find the amplitude equations to be:
\begin{align}
\dot{\Psi}_1 &= \lambda_1(q)\Psi_1+w\Psi_1{}^3+x\Psi_1\Psi_2{}^2, \label{eq:amplitudePsi1}\\
\dot{\Psi}_2 &= \lambda_2(q)\Psi_2+y\Psi_2{}^3+z\Psi_2\Psi_1{}^2,
\label{eq:amplitudePsi2}
\end{align}
where $w$, $x$, $y$, and $z$ are given by:
\begin{align}
w &= \left(\mathcal{Z}-\frac{\pi^2\mathcal{K} K}{ L^2}b_1(q)\right)\alpha_{4,0}+\left(\mathcal{H}+\mathcal{K}\tilde{h}_c\right)\beta_{4,0}, \\
x &= \left(3\mathcal{Z}-\frac{\pi^2\mathcal{K} K}{ L^2}b_1(q)-\frac{2\pi^2\mathcal{K} K}{ L^2}b_2(q)\right)\alpha_{2,2}+3\left(\mathcal{H}+\mathcal{K}\tilde{h}_c\right)\beta_{2,2}, \\
y &= \left(\mathcal{Z}-\frac{\pi^2\mathcal{K} K}{ L^2}b_2(q)\right)\alpha_{0,4}+\left(\mathcal{H}+\mathcal{K}\tilde{h}_c\right)\beta_{0,4},  \\
z &= \left(3\mathcal{Z}-\frac{\pi^2\mathcal{K} K}{ L^2}b_2(q)-\frac{2\pi^2\mathcal{K} K}{ L^2}b_1(q)\right)\alpha_{2,2}+3\left(\mathcal{H}+\mathcal{K}\tilde{h}_c\right)\beta_{2,2},
\end{align}
where the quantities $\mathcal{K}$, $\mathcal{H}$ and $\mathcal{Z}$ are defined in Eqs. (\ref{eq:Hdef},\ref{eq:Kdef},\ref{eq:Zdef}). We plot these quantities with the parameters of Fig.~2 in the main text, see Fig.~\ref{figS_AmpEq}(a). Interestingly, we see that for larger negative curvatures, which corresponds to $q<0$, they all approach each other in value: to this linear approximation of the flow functions, the flow patterns become dynamically indistinguishable. Nevertheless, the standard analysis we perform below gives a very satisfactory result for the lower tricritical point.
\vskip0.5cm
\paragraph*{Conditions for the existence of a tricritical point} We consider specific regions of the phase diagram and values of the coefficient $\lambda_2$
\begin{itemize}
    \item If $\lambda_2(q)\ll 0$, the amplitude equation for the second mode amplitude $\Psi_2$ will, if $y,z<0$, tell us that $\Psi_2=0$.
This, in turn, if $w<0$, implies that the transition will be continuous, since larger values of $\Psi_1$ are heavily suppressed by the cubic term proportional to $w$. 
    \item If $\lambda_2\approx 0$, then, at steady state for $\Psi_2$,
\begin{equation}
0=y\Psi_2^3+z\Psi_2\Psi_1^2 \implies \Psi_2 = \sqrt{-\frac{z\Psi_1^2}{y}}. 
\end{equation}
The latter equation can then be inserted into the amplitude equation  for the first mode, Eq. (\ref{eq:amplitudePsi1}):
\begin{equation}
\dot{\Psi}_1=\lambda_1(q)\Psi_1+\left(w-\frac{xz}{y}\right)\Psi_1^3. \label{eq:weff}
\end{equation}
Should the effective cubic $w_{\mathrm{eff}} = w- xz/y$ coefficient changes sign and becomes positive, then we will treat this as evidence for a tricritical point, where the transition changes from continuous to discontinuous.
\end{itemize}

\vskip0.5cm
\paragraph*{Results: location of the tricritical point} 
\begin{itemize}
    \item Figure~\ref{figS_AmpEq}(b) shows that for the parameters of Fig.~2 in the main text, the zero effective cubic coefficient $w-xz/y=0$, see Eq. (\ref{eq:weff}) does indeed cross the threshold expression $\lambda_1=0$ in $\zeta_s$--$C_0$ space, in a region where $\lambda_2 \approx 0$. Our analytical prediction ($\zeta_s\approx -0.75$, $C_0\approx -2.4$) is in quantitative agreement with the lower tricritical point found in our numerical analysis, see Figs.~\ref{figS_C0zeta}(a)~and~(b).
    \item Although this argument of mode-coupling is probably not valid for the upper tricritical point (since $\lambda_2$ is not close to $0$), could a change of sign of the straight cubic coefficient $w$ give a good approximation for the upper tricritical point? We plot the curve $w=0$ in Fig.~\ref{figS_AmpEq}(c), and see that the intersection with the threshold $\lambda_1 = 0$ is still a small distance from the numerically determined tricritical point, which is located directly at zero activity. To get a more accurate location still would require consider of even higher terms in the $\phi$-expansion.
\end{itemize}

\subsection{Linear stability analysis in the presence of substrate friction}
For non-zero friction, we write the linearized system~\eqref{eq:LinSys} using Fourier modes, $\delta\theta=\sum_m \delta\theta_m \exp[iq_m s]$, $\sigma_{sy}=\sum_m \sigma_{m}\exp[i q_m s]$, $v_y = \sum_m v_m\exp[iq_m s]$, with wavevectors $q_m=m\pi/L$ for all integers $m$.
Doing this, we may again reduce our system to one variable $\delta\theta$.
Since $C_{ss}(s)=C_0\cos ks$ is spatially varying, there is coupling between Fourier modes:
\begin{equation}
\dot{\delta\theta_m}=-\alpha_m\delta\theta_m+\beta_m(\delta\theta_{m+kL/2\pi}+\delta\theta_{m-kL/2\pi}),
\end{equation}
where
\begin{gather}
\alpha_m = \frac{q_m^2}{\xi+\eta q_m^2}\left[\frac{K\xi}{\gamma}+\frac{(\nu+1)\zeta_s}{2}+\frac{K\eta q_m^2}{\tilde{\eta}}\right],\\
\beta_m = -\frac{C_0}{2(\xi+\eta q_m^2)}\left[h_c\left(\frac{\xi}{\gamma}+\frac{\eta q_m^2}{\tilde{\eta}}\right)+\nu_c(\xi+\eta q_m^2)\right].
\end{gather}
The boundary conditions $\delta\theta(0)=\delta\theta(L)=0$ imply that the resulting Fourier series is based on sine functions ($\delta\theta_m=-\delta\theta_{-m}$), allowing us to consider only positive values of $m$.
To perform the stability analysis, we must diagonalise the resulting operator and find its eigenvalues.
We first note that since $k=2\pi/L$, the odd modes, $\delta\bm{\theta}^o = \{\,\delta\theta_{2m-1}\, \forall \, m \in \mathbb{Z}^{+} \,\}$, are completely decoupled from the even modes, $\delta\bm{\theta}^e = \{\,\delta\theta_{2m}\, \forall\, m \in \mathbb{Z}^{+} \,\}$, and so we may write them as two separate systems in the stability analysis:
\begin{gather}
\dot{\delta\bm{\theta}}^o =\begin{pmatrix}
-(\alpha_1 + \beta_1) & \beta_1 &  &  &   \\
\beta_3 & -\alpha_3 & \beta_3 &  &  \\
 & \ddots & \ddots & \ddots &    \\
 &  & \beta_{2m-1} & -\alpha_{2m-1} & \beta_{2m-1}  \\
 & & & \ddots & \ddots \\
\end{pmatrix} 
\delta\bm{\theta}^o, \label{eq:oddmat}  \\
\dot{\delta\bm{\theta}}^e =\begin{pmatrix}
-\alpha_2 & \beta_2 &  &  &   \\
\beta_4 & -\alpha_4 & \beta_4 &  &  \\
 & \ddots & \ddots & \ddots &    \\
 &  & \beta_{2m} & -\alpha_{2m} & \beta_{2m}  \\
 & & & \ddots & \ddots   \\
\end{pmatrix}\delta\bm{\theta}^e, \label{eq:evenmat}
\end{gather}
where $m$ is a positive integer.

The modes associated to the largest eigenvalues of the odd and even mode operators ($\lambda_1$ and $\lambda_2$ respectively), will in general have large components of the first odd and even Fourier modes respectively.
Therefore, the first odd flow mode to become unstable is a single shear state, and the first even flow mode to become unstable is a double shear state.
As for the non-friction case, finding the threshold expressions $\lambda_{1,2} = 0$ will allow us to construct the phase diagram.

We did not find an exact solution for the eigenvalues.
Instead, since the on-diagonal $\alpha_m$ terms grow quadratically with $m$, while the $\beta_m$ terms are constant at large $m$, we reason that the higher Fourier modes are very strongly suppressed, and so we can get a reasonable approximation to the true eigenvalues by truncating the operator. We truncate both the even and odd operators to $3\times3$ matrices, finding the largest eigenvalues for both.
Then we generate two threshold curves, $\lambda_1 = 0$ and $\lambda_2 = 0$, as in Fig.~\ref{fig:fric}(a,b). For both eigenvalues negative, $\lambda_1,\lambda_2<0$, there will be no flow. However, unlike the zero-friction case, $\lambda_1$ is not always larger than $\lambda_2$. Instead, the no-flow region is bordered by regions where \textit{either} the  odd (SS) pattern \textit{or} the even (DS) pattern is unstable. This is notably different to the non-friction case, where the no-flow state is bordered \textit{only} by the single shear state. Hence, the presence of friction permits a direct continuous NF--DS transition. The resulting phase diagram is constructed by taking the line segments from $\lambda_1=0$ and $\lambda_2=0$ that butt up directly against the no-flow region, see Fig.~\ref{fig:fric}(a)--(d), and agrees excellently with the numerical results.

The analysis we have performed here will not allow us to make assertions about pattern selection far from the no-flow state where both even and odd flow patterns have positive eigenvalues. The pattern selection of Fig. \ref{fig:fric}(c-f) is determined through the numerical simulations.

\newpage
\clearpage

\begin{figure}[h!]
\includegraphics[width=13.5cm]{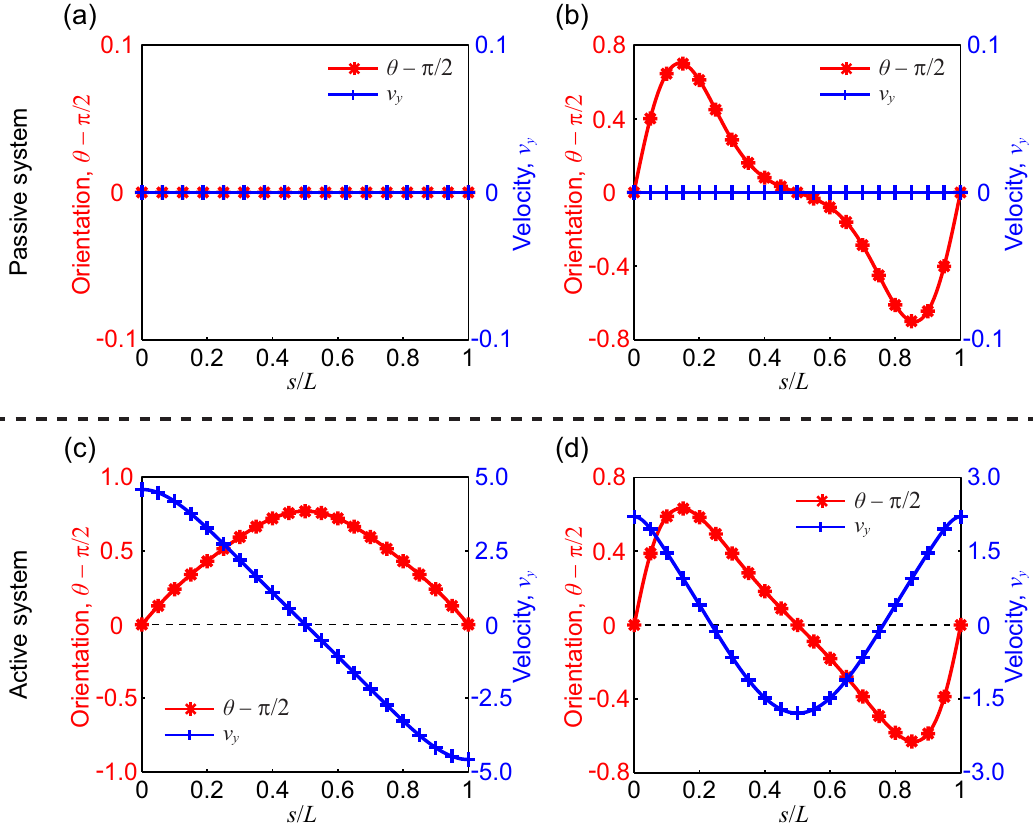}
\caption[Velocity and orientation profiles supporting Fig. 1 of the main text.]{Profiles of the velocity ${{v}_{y}}$ and the orientation $\theta $ for the typical patterns shown in Fig. 1(c–f) in the main text. Parameters: (a) $\xi =0$, ${{\nu }_{C}}=0$, ${{h}_{c}}=1$, ${{\lambda }_{s}}=0$, $L=10$, and ${{C}_{0}}=0.6$; (b) $\xi =0$, ${{\nu }_{C}}=0$, ${{h}_{c}}=1$, ${{\lambda }_{s}}=0$, $L=10$, and ${{C}_{0}}=-1$; (c) $\xi =0$, ${{\nu }_{C}}=-1$, ${{h}_{c}}=1$, ${{\lambda }_{s}}=4$, $L=10$, and ${{C}_{0}}=0.6$; (d) $\xi =0$, ${{\nu }_{C}}=-1$, ${{h}_{c}}=1$, ${{\lambda }_{s}}=4$, $L=10$, and ${{C}_{0}}=-1$.}
\end{figure}

\begin{figure}[h!]
\includegraphics[width=12.5cm]{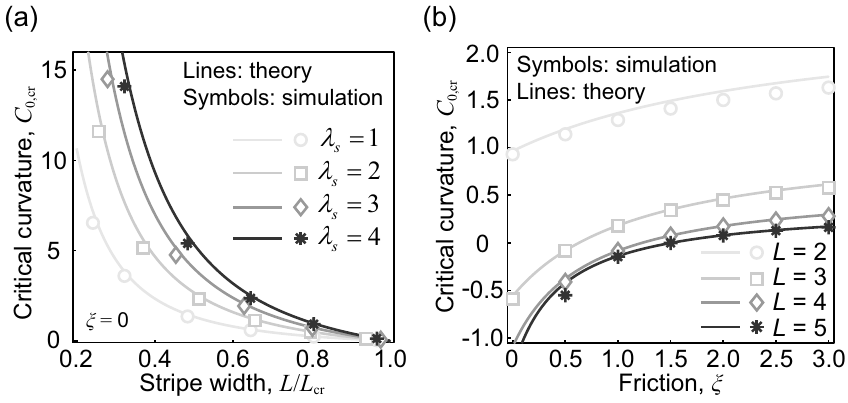}
\caption[Analytical vs numerical results for the critical curvature.]{Comparison of the predicted critical curvature ${{C}_{0,\text{cr}}}$ (see main text, Eqs. (6) and (7)) with numerical simulations. (a) The critical curvature ${{C}_{0,\text{cr}}}$ in the absence of friction ($\xi =0$) versus the rescaled stripe width ${L}/{{{L}_{\text{cr}}}}\;$ for various active splay traction ${{\lambda }_{s}}$. Here, ${{L}_{\text{cr}}}$ is the critical stripe width for flat geometry. (b) The critical curvature ${{C}_{0,\text{cr}}}$ versus the friction $\xi $ for various stripe width $L$.}
\end{figure}

\begin{figure}[h!]
\includegraphics[width=10.5cm]{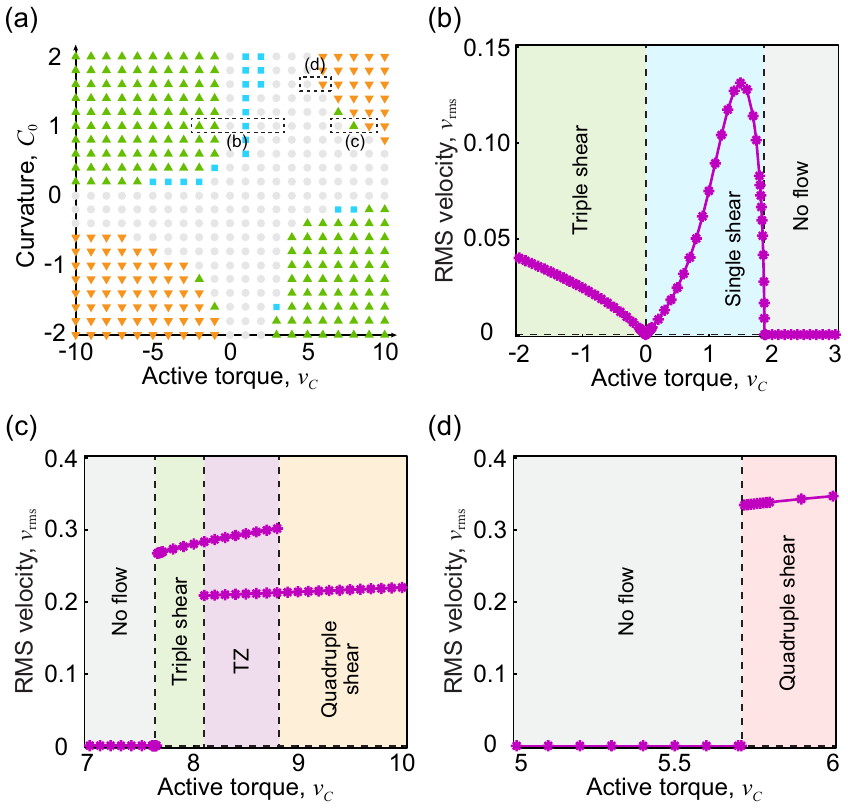}
\caption[The effect of $\nu_c$ on tissue flow patterns.]{Flows dictated by the curvature coupling $\nu_c$ and the curvature $C_0$. (a) Flow pattern as a function of the active torque $\nu_c$ and the curvature $C_0$: no flow (grey); single shear (blue); triple shear (light green)) and quadruple shear (light yellow). (b, c) Flows as a function of the active torque $\nu_c$, which corresponds to the dash boxes in (a) with $C_0=1.0$. (d) Flows as a function of the active torque $\nu_c$, which corresponds to the dash boxes in (a) with $C_0=1.6$. Parameters: $\xi=0$, $h_c = 2$, $\zeta = \lambda_s = \lambda_b = 0$, and $L=6$.}
\end{figure}

\newpage
\clearpage

\begin{figure}[h!]
\includegraphics[width=0.9\textwidth]{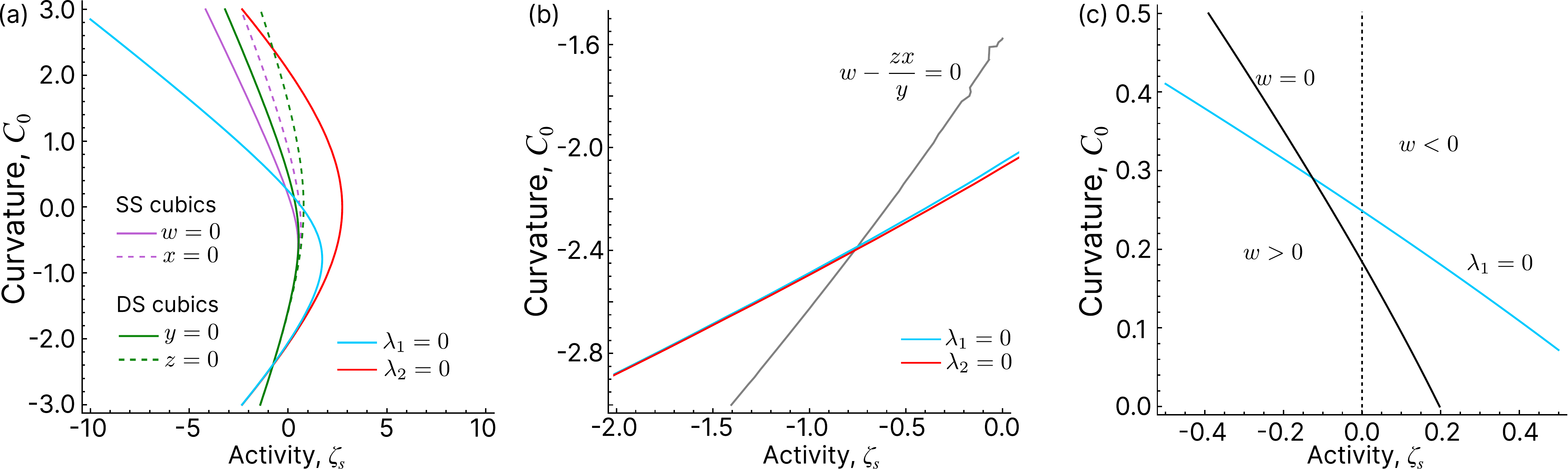}
\caption[Analytical plots of amplitude equation terms.]{Analytical plots of amplitude equation terms. (a) The eigenvalues for the SS mode ($\lambda_1$) and DS mode ($\lambda_2$) become degenerate for larger negative values of $C_0$. In fact, the cubic terms in the two amplitude equations also converge to one curve. (b) The change of sign of the effective cubic $w-xz/y$ when $\lambda_2\approx 0$ and $\lambda_1 = 0$ is some indication of a tricritical point in the vicinity (compare with Fig.~\ref{figS_C0zeta}(a) and (b)). (c) A change of sign of $w$ (the $\Psi_1^3$ coefficient, see Eq. (\ref{eq:amplitudePsi1})) does not pin down the exact location of the upper tricritical point to $\zeta_s=0$, as shown in simulations (Fig.~\ref{figS_C0zeta}(a) and (b)).}
\label{figS_AmpEq}
\end{figure}

\begin{figure}[b!]
\includegraphics[width=12.5cm]{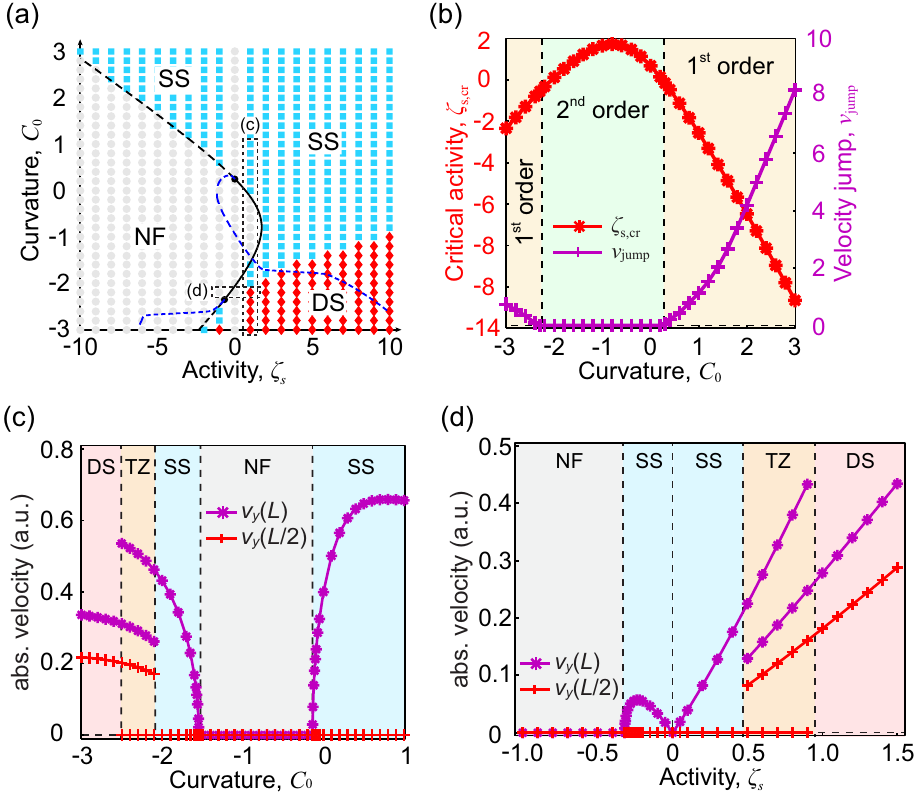}
\caption[Pattern selection dictated by $\zeta_s$ and $C_0$.]
{ Pattern selection dictated by $\zeta_s$ and $C_0$. 
(a) $\zeta_s$--$C_0$ flow pattern diagram and analytical prediction of the NF--SS transition (solid: continuous; dashed: discontinuous). 
The two black circular spots separate the continuous transition and the discontinuous transition.
The blue dashed curve represents the discontinuous SS--NF (or SS--SS) transition while increasing (or decreasing) $\zeta_s$ from a SS pattern.
(b) The critical activity $\zeta_{s,\rm cr}$ (corresponding to the black curve in (a)) and the velocity jump $v_{\rm jump}$ upon the NF--SS transition as a function of the curvature $C_0$. 
(c) Constant $\zeta_s=1$ cut through (a) shows that a sufficiently large curvature value of either signs leads to NF--SS transition.  
(d) Constant $C_0 = -2.2$ cut through (a). We can see a thresholdless flow transition at $\zeta_s = 0$: further increasing the activity $\zeta_s$ (in the positive regime), we observe a discontinuous SS--DS transition; while decreasing the activity $\zeta_s$, we observe a continuous SS--NF transition. 
Parameters: $\xi=0$, $h_c = 2$, $\nu_c = 0$, and $L=6$.}
\label{figS_C0zeta}
\end{figure}

\newpage
\clearpage

\begin{figure}[h!]
\includegraphics[width=17.5cm]{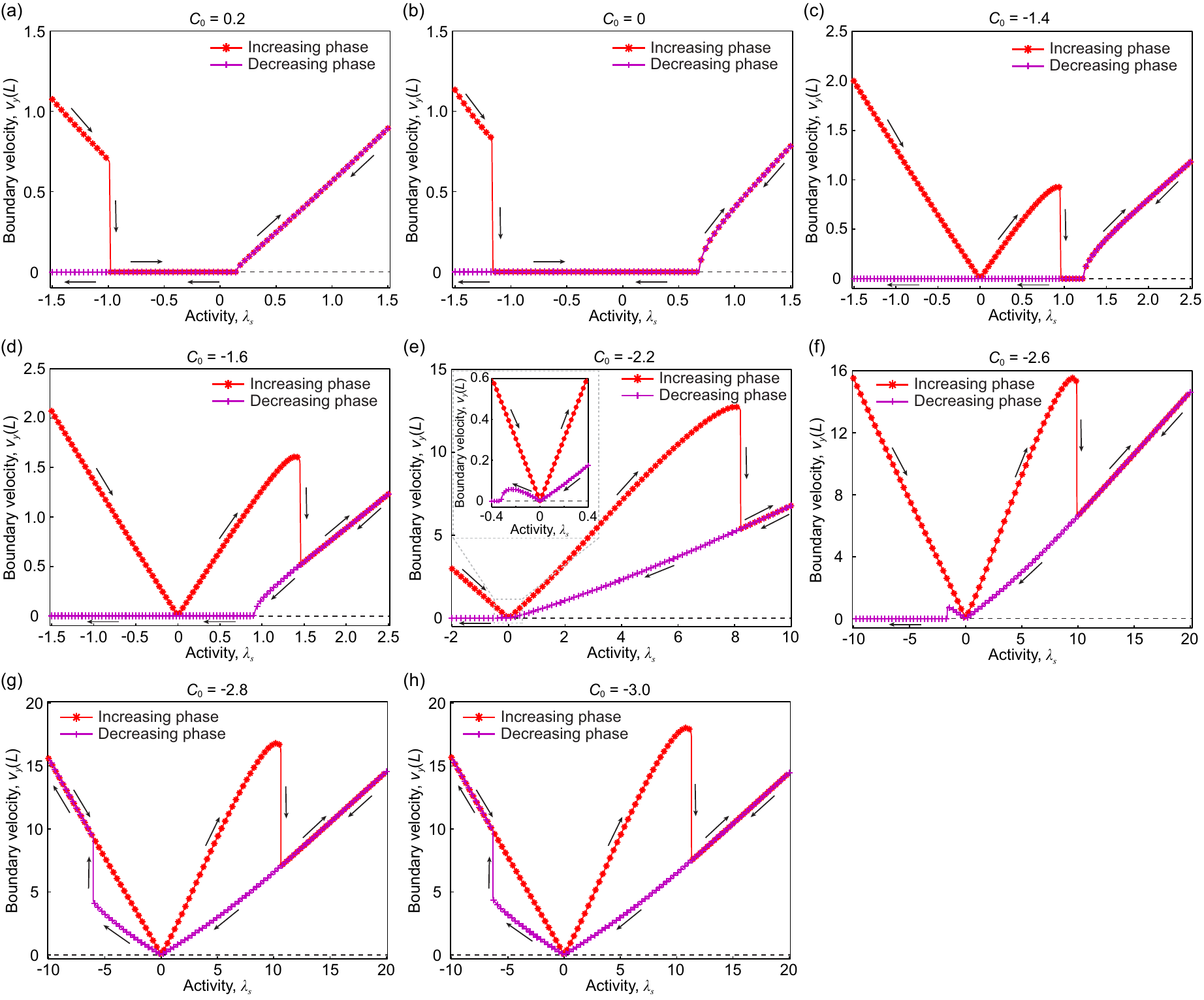}
\caption[Pattern selection upon quasi-static variation of splay activity $\zeta_s$.]{Pattern selection upon varying the splay activity $\lambda_s$ quasi-statically (see quenching simulation method in the SI Sec. \ref{sec:QuenchingSimulation}) for different curvatures: (a) $C_0 = 0.2$; (b) $C_0 = 0$; (c) $C_0 = -1.4$; (d) $C_0 = -1.6$; (e) $C_0 = -2.2$; (f) $C_0 = -2.6$; (g) $C_0 = -2.8$; (h) $C_0 = -3.0$. Here, we increase the splay activity $\lambda_s$ quasi-statically from negative values to positive values (the single shear regime); then decrease it quasi-statically back to negative values. The initial single shear flow patterns are obtained by quenching simulation of the curvature $C_0$ (e.g., see Fig. 2(f) in the main text). Shown here are the boundary velocity $v_y (L)$ as a function of the quasi-statically varying splay activity $\lambda_s$, where the black arrows indicate the route. Parameters: $\xi=0$, $h_c = 2$, $\nu_c = 0$, and $L=6$.}
\end{figure}

\newpage
\clearpage

\begin{figure}[h!]
\includegraphics{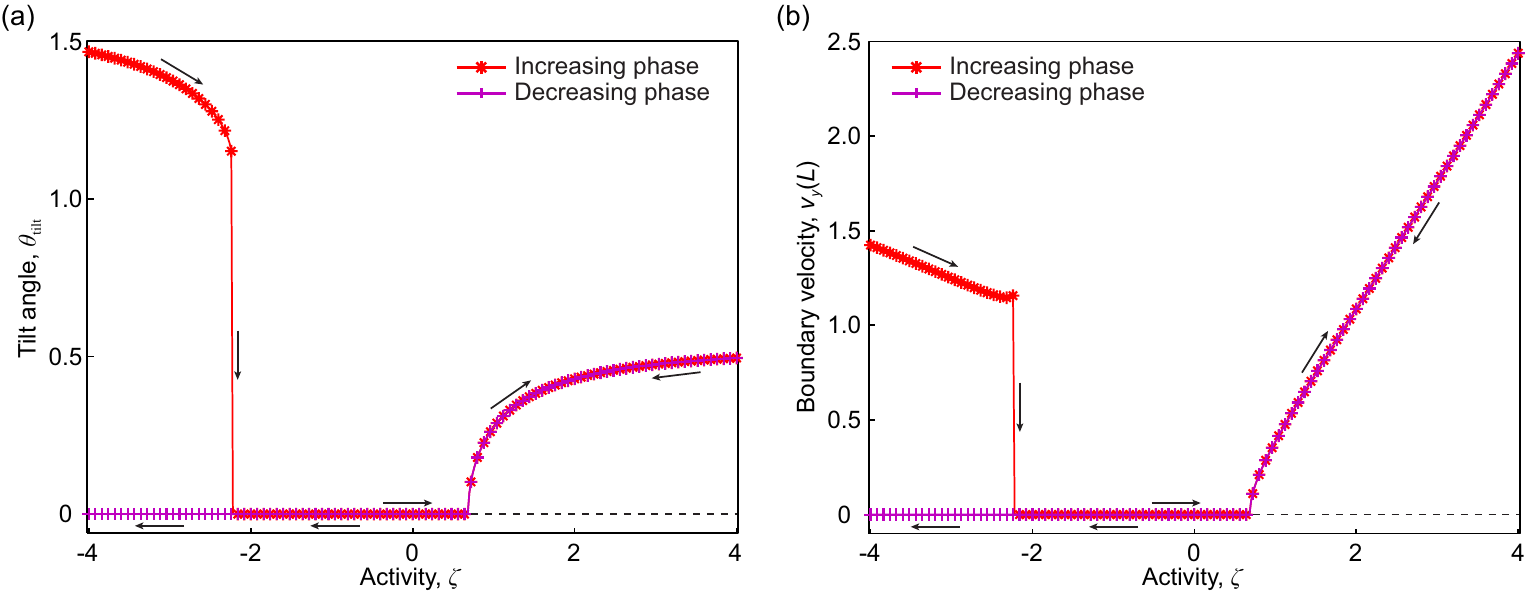}
\caption[Pattern selection upon quasi-static variation of bulk activity $\zeta$.]{Pattern selection upon varying the bulk activity $\zeta$ quasi-statically for a flat geometry (see the quenching simulation method in the SI Sec. \ref{sec:QuenchingSimulation}). Here, we increase the bulk activity $\zeta$ quasi-statically from $-4$ to $4$; then decrease it quasi-statically from $4$ to $-4$. The initial single shear flow pattern (at $\zeta = -4$) is obtained by quenching simulation of curvature $C_0$. (a) The tilt angle $\theta_{\rm tilt}$ and (b) the boundary velocity $v_y (L)$ as a function of the quasi-statically varying bulk activity $\zeta$, where the black arrows indicate the route. Parameters: $\xi=0$, $h_c = 2$, $\nu_c = 0$, $\lambda_s = \lambda_b = 0$, $C_0 = 0$, and $L=6$.}
\end{figure}

\begin{figure}[t!]
\centering
\includegraphics[width=16.7cm]{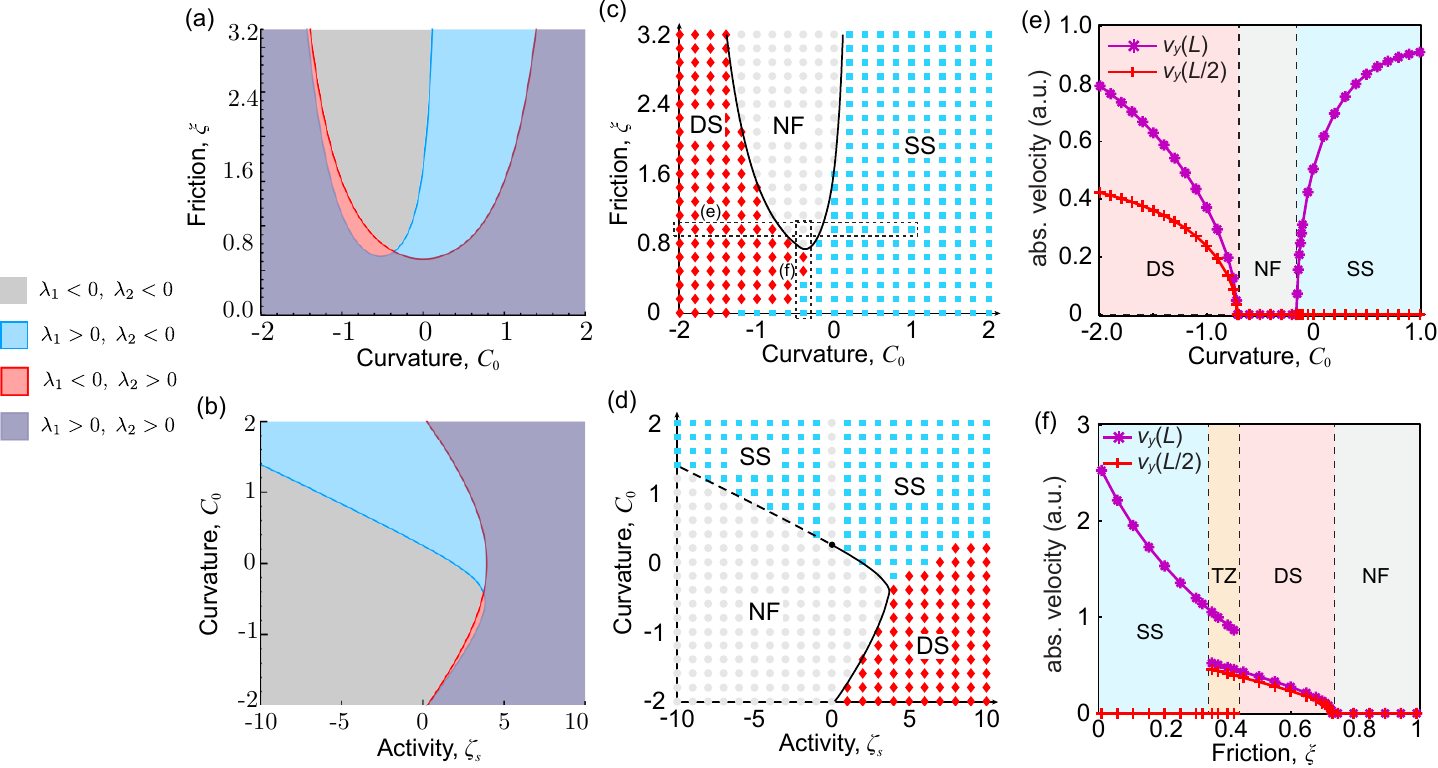}
\caption[Flows in the presence of friction.]{Flows in the presence of friction. (a), (b): Analytical curves for the least stable modes of the odd and even mode systems in Eqs.~\eqref{eq:oddmat}--~\eqref{eq:evenmat} for (a) fixed $\zeta_s=4$, and (b) fixed $\xi=0.6$, with the other parameters as in Table~\ref{table_s1}. The matrices have been truncated to $3\times 3$.
(c)--(f): Numerical results for (c) fixed $\zeta_s=4$, and (f) fixed $\xi=0.6$;
(c) Flow pattern as a function of the curvature $C_0$ and friction coefficient $\xi$: no flow (grey, NF); single shear (blue, SS); and double shear (red, DS). The black solid curve represents a continuous NF--SS or NF--DS transition. 
(d) Flows as a function of the activity $\zeta_s$ and curvature $C_0$. The black solid curve represents a continuous NF--SS or NF--DS transition; the black dashed curve refers to a discontinuous NF--SS transition. 
(e) Flows as a function of the curvature $C_0$ for $\xi=1$ (see horizontal dashed box in (a)). 
(f) Flows as a function of the friction $\xi$, which corresponds to the vertical dash box in (a) with $C_0=-0.4$. 
Parameters: $h_c=2$, $\nu_c = 0$, and $L=6$.}
\label{fig:fric}
\end{figure}


\begin{figure}[t!]
\includegraphics[width=17.5cm]{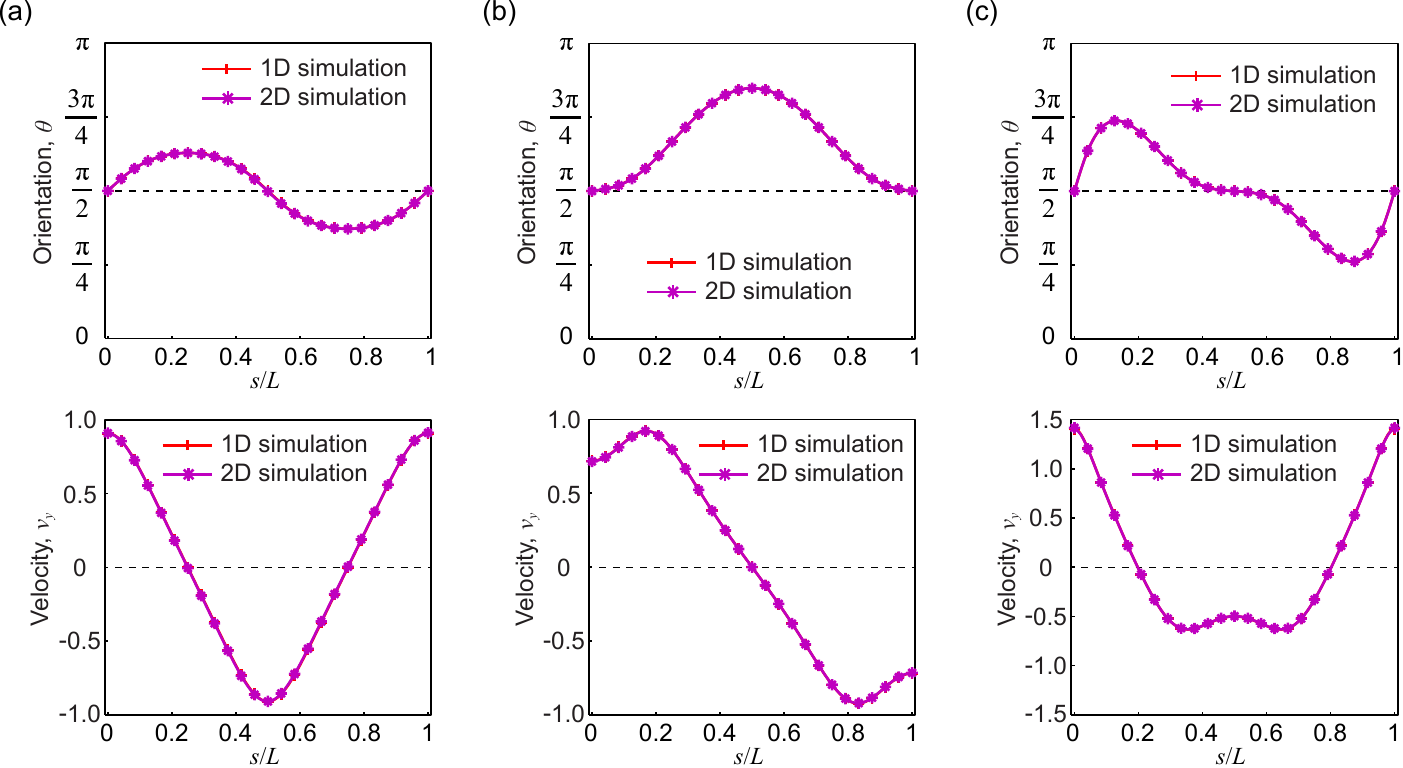}
\caption[Comparing 1D and 2D simulation results.]{Comparisons of the orientation profile and the velocity profile obtained from 1D simulations and 2D simulations. (a) $C_0 = 0$. (b) $C_0 = 0.5$. (c) $C_0 = -1$. Other parameters: $\xi =0.6$, ${{h}_{c}}=1$, ${{\lambda }_{s}}=4$, and $L = 12$.}
\end{figure}

\begin{figure}[t!]
\includegraphics[width=12cm]{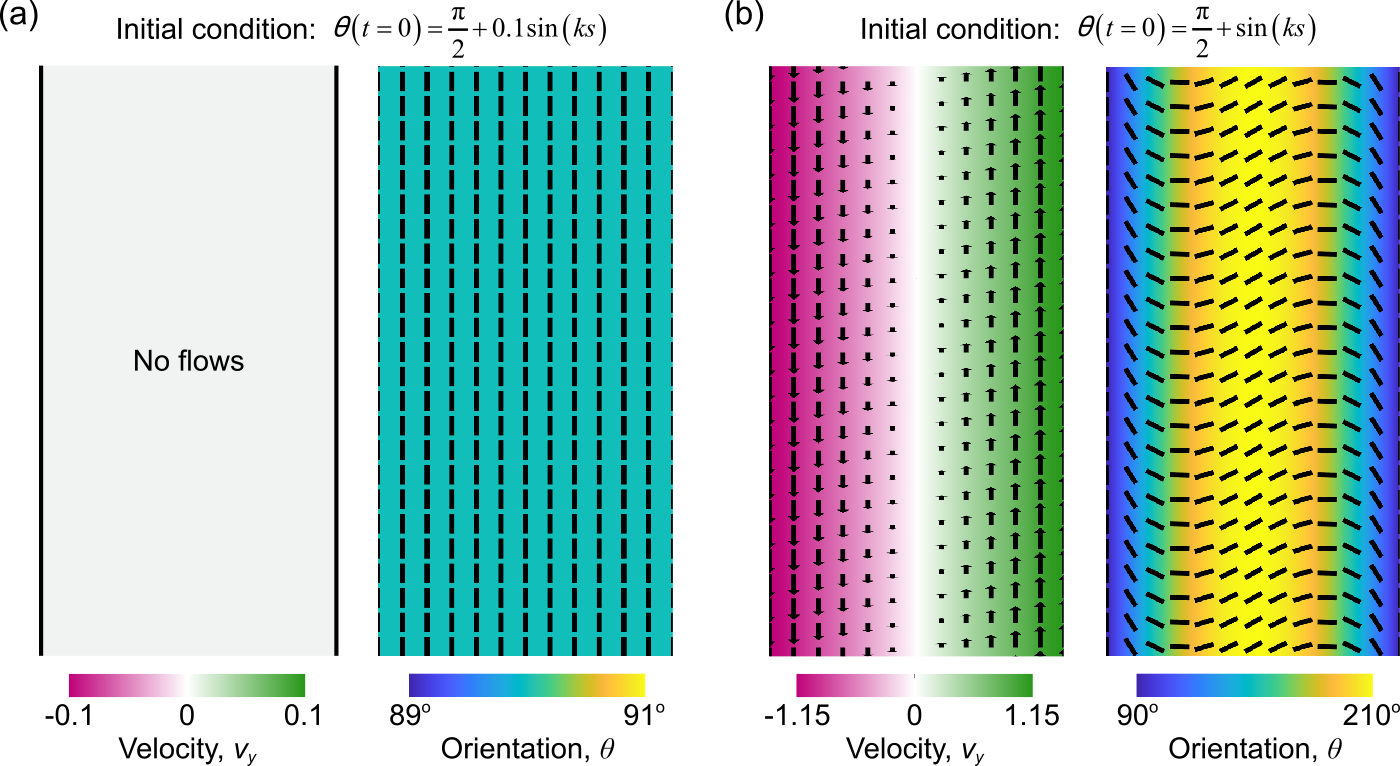}
\caption[Stable single shear flow from 2D simulations in the negative activity regime.]{Stable single shear flow pattern obtained in 2D simulations in the negative activity regime for flat geometries. Here we show two different patterns obtained by setting the initial orientation profile $\theta (t = 0) = \dfrac{\pi}{2} + A_{\theta} \sin \left( ks \right)$: (a) small perturbation from the uniform pattern with $A_{\theta} = 0.1$; (b) large perturbation from the uniform pattern with $A_{\theta} = 1.0$. Parameters: $\xi = 0$, $\zeta = \lambda_b = 0$, $\lambda_s=-1.5$, $L = 6$, and $C_0 = 0$.}
\end{figure}

\newpage
\clearpage

\begin{figure}[t!]
\includegraphics[width=17.5cm]{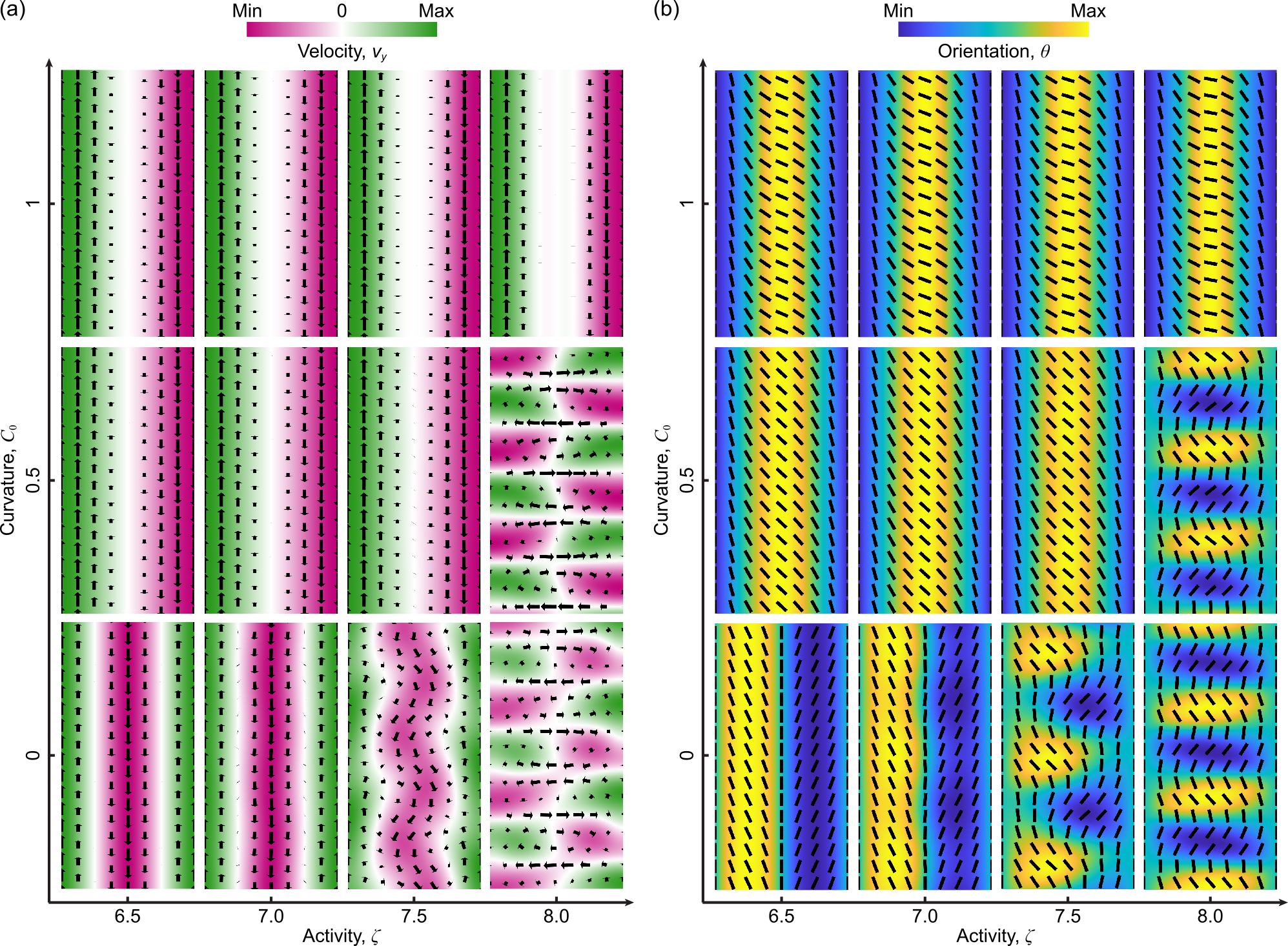}
\caption[Fixing $\zeta_s$ while varying the bulk contractility $\zeta$.]{Changing flow patterns upon varying the activity parameter $\zeta$ while keeping $\zeta_s = \zeta + \lambda_s$ fixed. Here we show the velocity field (a) and the orientation field (b) for different curvature case ($C_0 = 0, \ 0.5, \ 1.0$). In (a), the color code represents $v_y$ and arrows denote velocity vectors; in (b), the color code refers to $\theta$ and lines for orientation directors. For the flat case, we can clearly see a transition from the double shear flow pattern to the vortex chain pattern upon increasing the activity parameter $\zeta$. However, the curvature $C_0$ tends to destroy the vortex chain pattern and stabilize the shear flow pattern. Parameters: $\xi =0.6$, $h_c=1$, $\zeta_s=8$, and $L = 8$.}
\end{figure}

\newpage
\clearpage

\begin{table}[h!]
\centering
\caption[Table of parameter values]{List of default parameter values used in simulations.}\label{table_s1}
\begin{threeparttable}
{\def\arraystretch{1.5}
\begin{tabular}{p{3cm}<{\centering} p{6cm}<{\centering} p{2cm}<{\centering}}
\toprule[1.0pt]
Parameter & Description & Value  \\ \midrule[0.5pt]
$\xi$ & Substrate friction & 0.6  \\ 
$\eta$ & Tissue shear viscosity & 1  \\ 
$\gamma$ & Rotational viscosity & 1 \\ 
$\nu$ & Flow alignment parameter & $-2$  \\ 
$\nu_c$ & Active curvature coupling & $0$   \\ 
$K_1 = K_3 = K$ & Nematic elasticity & 1   \\ 
$\zeta$ & Active stress parameter & 0  \\ 
$\lambda_s$ & Active splay traction & 4 \\ 
$\lambda_b$ & Active bend traction & 0 \\ 
\bottomrule[1.0pt]
\end{tabular}
}
\end{threeparttable}
\end{table}